\documentclass[preprint,10pt]{aastex}

\newcommand{\B}{\phn{ }}
\newcommand{\scr}{\scriptsize}
\newcommand{\lan}{\langle}
\newcommand{\ran}{\rangle}
\newcommand{\asec}{\hbox to 1pt{}\rlap{$^{\prime\prime}$}.\hbox to 2pt{}}

\shorttitle{\emph{HST} Morphologies of E+A's}
\shortauthors{Yang et al.}

\begin{document}
\title{E+A Galaxies and the Formation of Early Type Galaxies at z$\sim$0 \altaffilmark{1}}

\author{Yujin Yang, Ann I. Zabludoff, and Dennis Zaritsky}
\affil{Steward Observatory, University of Arizona, Tucson, AZ 85721}
\email{yyang@as.arizona.edu, azabludoff@as.arizona.edu, dennis@as.arizona.edu}

\author{Tod R. Lauer}
\affil{National Optical Astronomy Observatories, Tucson, AZ 85726}
\email{lauer@noao.edu}

\author{J. Christopher Mihos\altaffilmark{2}}
\affil{Department of Astronomy, Case Western Reserve University, 10900
  Euclid Ave, Cleveland, OH 44106}
\email{hos@burro.astr.cwru.edu}

\altaffiltext{1} {Based on observations with the NASA/ESA Hubble Space
  Telescope obtained at the Space Telescope Science Institute, which
  is operated by the Association of Universities for Research in
  Astronomy, Incorporated, under NASA contract NAS5-26555.}
\altaffiltext{2} {NSF CAREER Fellow and Research Corporation Cottrell Scholar}

\begin{abstract}
E+A galaxies, whose spectra have deep Balmer absorption lines 
but no significant [OII] emission, are the best candidates 
for an evolutionary link between star-forming, gas-rich galaxies 
and quiescent, gas-poor galaxies. 
Yet their current \emph{morphologies} are not well known. 
We present \emph{HST}/WFPC2 observations of the five bluest E+A galaxies
($z\sim0.1$) in the Zabludoff et al. sample to study 
whether their detailed morphologies are consistent with late-to-early type 
evolution and to determine what drives that evolution.
The morphologies of four galaxies are disturbed, indicating that
a galaxy-galaxy merger is at least one mechanism that leads to the E+A phase.

Two-dimensional image fitting shows 
that the E+As are generally bulge-dominated systems, 
even though at least two E+As may have underlying disks.
In the Fundamental Plane, E+As stand apart from the E/S0s 
mainly due to their high effective surface brightness.
Fading of the young stellar population and the corresponding increase 
in their effective radii will cause these galaxies to migrate 
toward the locus of E/S0s.
E+As have profiles qualitatively like those of normal power-law 
early-type galaxies, but have higher surface brightnesses.
This result provides the first direct evidence supporting the hypothesis 
that power-law ellipticals form via gas-rich mergers.
In total, at least four E+As are morphologically consistent with 
early-type galaxies.

We detect compact sources, possibly young star clusters, 
associated with the galaxies.
These sources are much brighter ($M_R\sim-13$) than Galactic globular 
clusters, have luminosities consistent with the brightest clusters 
in nearby starburst galaxies, and have blue colors consistent with 
the ages estimated from the E+A galaxy spectra (several $10^8$ yr). 
Further study of such young star cluster candidates might provide 
the elusive chronometer needed to break the age/burst-strength degeneracy 
for these post-merger galaxies.
\end{abstract}


\section{Introduction}

If galaxies evolve morphologically from late to early types, then
some may be now changing from star-forming, gas-rich, disk-dominated
objects into quiescent, gas-poor spheroidals.  Spectroscopic surveys
have identified at least one set of candidates for such a
transformation: ``E+A'' galaxies\footnote{
Because their spectra are a superposition of a young stellar 
population (represented by A stars) and an old population 
(characterized by K stars), these galaxies became known as 
``E (for elliptical) + A'' galaxies \citep{Dressler} or,  
more straightforwardly, ``K+A'' or ``k+a'' galaxies
\citep{Franx,Dressler99,Poggianti}.},
whose spectra have deep Balmer absorption lines but no significant 
[OII] emission,
indicating that star formation ceased abruptly in these galaxies
within the last $\sim$ Gyr.  In general, E+A galaxies lack significant
amounts of HI gas \citep{chang} and have hot, pressure-supported
kinematics \citep{norton}, suggesting that these galaxies are indeed
evolving --- somehow --- from late to early types.  However, we do not
yet know whether their current {\it morphologies} are consistent with
late-to-early type evolution or what drives E+A evolution.

While the mechanism (or mechanisms) that causes galaxies to pass
through an E+A phase is not understood, there are several clues.
First, E+A spectra suggest a recent burst of star formation that
required the rapid consumption or dispersal of a gas reservoir.  Second,
although they were first studied in distant clusters \citep{Dressler}, 
E+As --- at least at low redshifts ($z\sim0.1$) --- lie mostly in low density
environments \citep{zab96,hogg}.  Third, in low-resolution POSS images, 
some E+As have
features suggestive of tidal tails \citep{zab96}.  Could E+As be the
result of disk galaxy mergers, which are both common in the field and
known to enhance star formation?  In the merger hypothesis, E+As are
further along the ``Toomre sequence'' \citep{toomre} and thus more
relaxed than systems like the Antennae, whose morphology and
kinematics are in such disarray that it is nearly impossible to
constrain its endproduct.  E+As may thus teach us considerably more
about the endpoints of galaxy-galaxy mergers.

We cannot test this picture of E+A formation, or whether the E+A phase
is a bona fide late-to-early type transition, without detailed
morphological information.  Simulations predict that well-evolved
major mergers have a hybrid morphology, including fading, low surface
brightness tidal tails at large radii, a more relaxed
spheroid-dominated core, and a population of young star clusters 
\citep{Barnes,Barnes_Hernquist,Ashman_Zepf,Mihos}.
Identifying such low surface brightness or small scale features, even
at low redshifts, requires spatial resolution on the order of 100 pc and
low sky background levels.  Therefore, Hubble Space Telescope imaging
of nearby E+As is required.

% plan of the paper
In this paper, we present the detailed \emph{HST}/WFPC2 morphologies
of the five bluest E+A galaxies in the \citet{zab96} sample.  We
review the sample and the data reduction methods in \S\ref{sec:reduction}.
We describe the qualitative morphologies of these galaxies in
\S\ref{sec:impression}, 
discussing the observed tidal features and the implications for E+A
origin.  We address the question of whether E+As are consistent with
evolution into early types by fitting two-dimensional, surface
brightness models to each image and deriving structural
parameters such as bulge-to-disk ratio, effective radius, and central
surface density (\S\ref{sec:fit}).  
In \S\ref{sec:color}, 
we examine the color gradients in the E+As
and compare them with the expectations from disk merger models.  
We compare the results with the fundamental plane for early type galaxies 
and with the surface brightness profiles of the nearby elliptical galaxies 
in \S\ref{sec:FP} and \S\ref{sec:nuker}, respectively.  
In \S\ref{sec:compact},
we search for star clusters in the E+As and ask whether their
properties are consistent with late-to-early type galaxy evolution.
We discuss the implications of our results for higher redshift galaxy
surveys in \S\ref{sec:highz}, cautioning that bulge-to-disk decompositions,
quantitative measures of asymmetry, and tests to uncover tidal
features may mislead.  Section \ref{summary} summarizes our results.

\section{Observations and Data Reduction}
\label{sec:reduction} 
Our \emph{HST} imaging sample is a subset of the 20 nearby
E+A galaxies\footnote{One (EA20) of the original 21 galaxies 
turned out to be misclassified as an E+A due to noise 
in the region of one of the spectral line diagnostics \citep{norton}.}
that were spectroscopically identified from 11,113 galaxy
spectra in the Las Campanas Redshift Survey (LCRS) with redshifts between
0.07 and 0.18 \citep{zab96}.  These E+As are selected by requiring
that their spectra have strong Balmer absorption features (average
equivalent width $\lan H \ran$ of H$\beta$, H$\gamma$ and H$\delta$ 
$>$ 5.5\AA) and little if any [OII] emission (EW[OII] $<$ 2.5 \AA).
Three-quarters of the E+As in the sample are in the field, 
well outside rich cluster environments.
The number of each E+A (e.g., EA1) is from \citet{zab96} and increases 
with increasing 4000{\AA} break ($D_{4000}$) strength. 
$D_{4000}$ is related to the galaxy's color 
--- bluer galaxies have smaller $D_{4000}$.
EA1 through EA5, the focus of our \emph{HST} study, 
have the smallest $D_{4000}$'s,
and therefore are more dominated by the young stellar population than 
the other E+As. This dominance can arise either because they
have had the most recent or strongest bursts. 
For the remainder of this paper, we refer to each galaxy by its 
assigned number.
Table \ref{tab:basic} summarizes the basic data of five galaxies: 
coordinates, redshift, and environment. Throughout this paper, 
we assume $H_0=70\ \mathrm{km\ s^{-1}\ Mpc^{-1}}$, $\Omega_\mathrm{M}=0.3$, 
and $\Omega_{\Lambda}=0.7$.

We obtained high resolution images of the five nearby ($z\sim0.08-0.12$) 
E+A galaxies with the
\emph{Hubble Space Telescope} Wide Field Planetary Camera 2 (WFPC2).
Because our sample is at relatively low redshift
(typically $z\sim0.1$, for which $0.5''$= $\sim1$ kpc), it is 
possible to study the LCRS sample in ways that are not
possible for the more classic E+As 
discovered in distant clusters. We take advantage of this
benefit to obtain spatially-resolved spectroscopy \citep{norton} and 
sub-kpc imaging here. 
We observed the sample using the
F702W ($\lambda_{\mathrm{eff}}=$ 6997\AA ) and 
F439W ($\lambda_{\mathrm{eff}}=$ 4292\AA ) filters and obtained
three CR-split 700s exposures for each object.
Stacked images were generated by summing the three individual images
for each galaxy and filter.  The pointing was identical for each image, 
so no shifting or interpolation was required.  We rejected CR events
by comparing deviant pixels within the stack to a WFPC2
noise model. 

We adopt photometric zero points of the \emph{HST}/WFPC2 from
\citet{Holtz} after correcting for the gain=7.0 and the nominal
infinite aperture. Our values are the  same as 
given in the \emph{HST Data Handbook}. 
For the Planetary Camera, F439W and F702W magnitude zero points 
are 20.884 and 22.428, respectively.
We adopt Galactic extinction corrections from \citet{Schl}, 
assuming an $R_V = 3.1$ extinction curve. $A_{\mathrm{F702W}}$ and
$A_{\mathrm{F439W}}$ are calculated from the relative extinction table
in the Appendix \citep{Schl}.  The value for F439W is not
available in the Appendix, so we use the extinction appropriate 
for the Landolt B magnitude.

To compare the magnitudes of galaxies within certain filters across a range
of redshifts, or to photometric models, we apply K-corrections.
In principle, the K-correction can be calculated by using the spectral
energy distribution (SED) of an object with full spectral
coverage and high S/N. Unfortunately, flux-calibrated spectra with full
spectral coverage and high S/N are not available for our sample. 
The SEDs of E+As strongly depend both on the stellar mass formed during
the starburst and on the time elapsed since the burst. To account for
this variation in stellar populations, we examine both extremes ---
a pure A type and a pure K type stellar spectrum.
We use A dwarf and K giant templates from the Gunn-Stryker spectrophotometric 
atlas \citep{gunnstyker},
which covers the wavelength range 3130 to 10800 \AA. 
We artificially redshift the template spectra to $(1+z)$ and 
measure the magnitude differences in the F702W and F439W filters using
the CALCPHOT routine within the IRAF/SYNPHOT package. 
In the F702W band the difference between the corrections for the
two populations is within $\sim$ 0.19 $-$ 0.31 magnitudes. In contrast, the
difference in the corrections is larger than 0.62 magnitudes for
F439W because the F439W 
filter band includes the Balmer jump. A slight shift of
the spectra can cause a large change in measured brightness.
We list both sets of corrections in Table \ref{tab:kcorr}, but
adopt the correction calculated for an A star with the justification
that these are the bluest, most A-like, of the E+As in the \citet{zab96} sample.
Because the K correction is the major source of uncertainty in 
our error budget, the global photometric quantities, especially colors,
possibly harbor significant systematic errors.  
The sense of any relative colors within 
a galaxy is not affected, although the numerical values may be.

\section{Results and Discussion}

The \emph{HST} images provide a wealth of information on the small and
large scale structure of these galaxies. With the goal of understanding the
origin of the E+A phenomenon and into what these systems will evolve, we
investigate the morphologies of these systems, their color profiles,
their location on the Fundamental Plane \citep{Jor96} of elliptical galaxies, and 
their relationship to ``core'' and ``power-law'' ellipticals \citep{f97}.
We also discover a population of associated point sources 
(possibly young star clusters).
Finally, we review 
the implications of our results, obtained for low-redshift E+As, for the
identification and study of such systems at higher redshifts.
The reader is referred to Tables 2-5 for a summary of the quantitative results
discussed in this section.

\subsection{Morphologies: First Impressions}
\label{sec:impression}
Figure \ref{fig:tile} shows the WFPC2 mosaic and PC images of our five E+A 
galaxies at different contrast levels. The full mosaic
images ($80''\times80''$) are in the left column. The center of each
E+A is located in the PC, which is in the upper right corner of each mosaic. 
Tidal features that extend into the other CCDs are evident in EA1-3.
The middle and right columns contain the F702W ($24''\times24''$) and 
F439W ($12''\times12''$) PC images, respectively, on a logarithmic flux scale. 
\emph{HST}/WFPC2 observations are relatively insensitive in the bluer band 
so that the signal in the F439W images typically extends out only to $\sim3$ kpc,
25\% of the red coverage, and even there it is of low signal-to-noise.

These five E+As exhibit a variety of morphologies ranging from 
a highly complex system (EA1) to what could visually be classified as
a barred S0 galaxy (EA5), 
even though they have been uniformly selected using spectroscopic 
criteria, i.e., ``k+a'' type spectra from the LCRS.

EA1 stands apart from the other four E+As.
It is composed of two components that are separated spatially 
by $\sim$ 3 kpc and another companion with a projected
separation of  14 kpc (assuming the companion is at the redshift of EA1).
The association is supported by an asymmetric feature emanating 
from the companion that could be tidal material
and a faint bridge that appears to connect it to EA1.

EA2 and EA3 also exhibit highly disturbed morphologies,
although EA3 could be visually classified as a normal face-on spiral galaxy 
in the low contrast PC image. This ambiguity in visual classification is
discussed in more detail in \S\ref{sec:highz}.
EA2 has tidal tail that extends to at least 50 kpc.

EA4 and EA5 appear less disturbed,
although EA4 has somewhat irregular outer isophotes, some lopsidedness 
(in the F439W filter image), and shell-like structures
closer to the center that are visible in the PC image.
The mechanism or mechanisms responsible for the spectral E+A phenomenon 
produce a variety of morphologies. Whether all of these systems will evolve
into a somewhat more homogeneous population --- for example, early-type galaxies --- 
is yet unclear.

\subsection{Morphologies : Bulge-Disk Decompositions}
\label{sec:fit}
While EA2-5 appear to have significant 
spheroidal components, EA3 and EA4, at least, also seem to have a flattened,
or perhaps disk-like, morphology.
Understanding the fate of these systems requires a quantitative estimate
of the relative importance of the dynamically hot and cold stellar components.

Measuring the surface brightness profile for asymmetric, disturbed
systems is challenging. To mitigate potential systematic problems,
we use two different algorithms.
First, to obtain photometric parameters, $r_e$ and $\mu_e$, we use
the two-dimensional image fitting algorithm GALFIT \citep{chien02}
designed to extract structural parameters directly from the galaxy
image. GALFIT assumes a two-dimensional model profile for the 
galaxy. The functional form of the models we choose to fit
include combinations of an $r^{1/4}$-law, a S\'ersic $r^{1/n}$-law, 
an exponential disk profile, and  a spatially constant sky background. 
We fit the following: the $(x,y)$ position of the center, 
$\mathrm{M}_{tot}$ (the total magnitude of the component), 
$r_e$ (the effective radius),  
$n$ (the S\'ersic index), $q$ (the axis ratio defined as $b/a$), 
the major axis position angle, and 
$c$ (the diskiness/boxiness index, where $c > 0$ indicates boxy). 
This index $c$ plays the same role as the $\cos 4\theta$
Fourier coefficient term used often in isophote analysis \citep{RZ}.  
As GALFIT explores parameter space, it convolves the model image with
a point-spread function (PSF) and compares it to the data for each
parameter set. 
The model PSFs are generated for each galaxy by the TinyTim
\citep{Krist99} software for the WFPC2.
Although convolution is computer intensive, the advantage of 
the convolution process is that it preserves the noise 
characteristics of the images and can be applied to low 
signal-to-noise images.

Because GALFIT begins with a very specific, smooth model, which
may be a poor representation of such distorted galaxies,
we also measure surface brightness
profiles using the IRAF/ELLIPSE algorithm. This approach allows the
center, major axis position angle,  and ellipticity of each ellipse to change, 
but does not enforce a model radial profile. 
To accurately recover the surface brightness profiles
without recourse to {\it ad hoc} models, we applied 20 iterations of Richardson-Lucy
deconvolution \citep{rich, lucy}. \citet{l98} showed
that the WFPC2 PSF can depress the brightness profile as far out as
$0\asec5$ from the galaxy center.  Richardson-Lucy deconvolution allows
the intrinsic brightness profile to be recovered to the few percent
level down to $r\sim0\asec05,$ with adequate exposure levels ($S/N\sim50$
in the galaxy center).  With reduced $S/N$ and only 20 deconvolution cycles,
the central ($r=0$) point in the profile may remain slightly-depressed,
dependent on the (unknown) intrinsic structure of the galaxy center.

Because EA1 is too disturbed to be reasonably
modeled by a simple disk+bulge model,
we restrict our analysis to EA2-5. For each galaxy,
we fit  three different light distributions: 
$r^{1/4}$ law, $r^{1/n}$ S\'ersic law, and $r^{1/4}$ + exponential disk law.
For EA2, we do not fit the $r^{1/4}$ + exponential disk law model
because we might be seeing this galaxy close to edge-on 
(see the linear residuals in Figure \ref{fig:residual}),
and it is hard for GALFIT to fit an edge-on disk with an extended tail.
The structural parameters and the reduced $\chi_\nu^2$'s of these three GALFIT models 
are listed in Tables \ref{tab:param} and 
\ref{tab:decomp}. With the exception of one case, $1 < \chi_\nu^2 < 2$. 
These values of $\chi_\nu^2$ are somewhat larger than statistically acceptable,
due presumably to the presence of asymmetric components, as can be seen
in Figure \ref{fig:residual}.

Of the three profiles we consider, only the S\'ersic profile has the
flexibility to model either a spheroidal or disk-like system by varying
the parameter $n$. Therefore the best-fit value of $n$ can guide our
conclusions about the nature of the galaxy. An exponential disk corresponds
to a value of $n = 1$, while the classic de Vaucouleurs profile corresponds to
$n = 4$. However, the correspondence between disk system, spheroid, and $n$ is
not quite this simple --- fitting S\'ersic profiles to SDSS galaxies, Blanton
et al. (2003) show a peak at $n = 1$ corresponding to disky systems, but no peak
at $n = 4$. Instead, spheroidal systems show a range of $n$ values. This result
is further complicated when one factors in differences in radial ranges
fit --- for example, fitting the inner slope of cuspy power-law ellipticals
(e.g., Lauer et al. 1995) will give a much higher $n$ value than will fits at
larger radii.

With these caveats in mind, we find that a single S\'ersic profile fit
yields $n > 5$ for all our galaxies, demonstrating that the light is dominated
by a spheroidal component. Indeed, the high values for $n$ indicate a
very high concentration of the light, even more than expected for a
classic de Vaucouleurs profile. Such high concentrations are consistent
with the idea that central starbursts have raised the central luminosity
density (e.g., Mihos \& Hernquist 1994). For example, in the case of EA4,
masking the inner kpc and refitting the Sersic law results in a value of
$n = 3.6$, much more typical of a normal elliptical. This is not always the
case, however --- in EA3, the high S\'ersic value persists even when the
nucleus is masked out. For EA3 the fitted value ($n = 8.7$) is unusually high
compared to normal ellipticals (e.g., Kelson et al. 2000; Graham et al. 2001;  
Graham 2002).  An additional complication to the interpretation of these
fits is that, while the light appears to be dominated by a concentrated
spheroid, some of the galaxies appear to contain an additional
disk-like component that would affect any dynamical model of a merger
and its aftermath.

To determine whether these galaxies do indeed contain a disk component,
we also fit models with two components.
The resulting radial profiles for EA3 and 4 (Figure \ref{fig:profile})
and the significant decrease in $\chi^2_\nu$ (an improvement in the fit at the
99\% confidence level) demonstrate that a pure spheroid model is not the
preferred model for these two systems. 
To avoid the degeneracies present in fitting disk and bulge simultaneously,
we also fit single S\'ersic profiles just to the outer parts of
the galaxies. Using the effective radii of the bulges calculated from 
the two-component fit, we mask pixels inside a chosen radius, vary that
radius to be $\sim$ $3-5 r_e$ and refit a single component.
We mark the effective radii and disk scale lengths with circles in
Figure  \ref{fig:residual}. 
For EA3, the best-fit S\'ersic indicies are $n=$ 2.5, 1.6 and 1.3  for
masks corresponding to $3 r_e$, $4 r_e$, and $5 r_e$, respectively. 
For EA4, we measure  $n=$ 0.93 and 0.86 
when we apply $3 r_e$ and  $4 r_e$ masks, respectively. In both of these cases, 
the S\'ersic index beyond several $r_e$ is as expected for 
an exponential disk and the fit spans 4 to 5 disk scale lengths.  
Although we cannot discriminate tidal material from a possible
underlying disk, 
we conclude that in EA3 and EA4 there is material beyond that
described by a spheroid and that it is consistent with an
underlying disk.
EA3 and EA4 appear to be sufficiently relaxed that no
significant dynamical evolution is expected, so they may become S0's.
We also hypothesize that their progenitors may have included a disk 
that was significantly heated but not completely destroyed 
during an intermediate mass ratio merger (e.g., Naab 2000; Bendo \& Barnes 2000).

Even though no disturbed tidal structure is apparent in EA5,
the modeling is complicated by the presence of a strong bar-like structure.
The presence of a bar-like feature suggests an underlying disk.
When viewed at the different contrasts in Figure \ref{fig:ea5}, 
EA5 is composed of at least three distinct components, 
a extended light distribution in outer part (axis ratio $q \sim 0.8$),
a compact and elliptical bar structure ($q \sim 0.4 - 0.5$),
and a very bright blue central nucleus.
The three-component fit gives the S\'ersic index $n=1.1$ for the central nucleus,
$n=0.5$ (Gaussian) for the bar, and $n=1.5$ for the outer disk-like region.
This three-component S\'ersic profile fit (Figure \ref{fig:ea5}) is acceptable and
suggests the presence of disk.
For mask sizes $2.5 r_e$, $3 r_e$ 
and $3.5 r_e$, the best fit S\'ersic indicies $n$ are 2.0, 1.8, and 1.8, respectively.
However, unlike for EA3 and EA4, we do this fit in a limited region and
cannot conclude that EA5 has a distinct exponential disk component.
The presence of a bar-like feature also suggests an underlying disk.

For the E+As that may contain a disk component,
we calculate a bulge-to-total light ratio (B/T) to quantify the
relative importance of the bulge and disk-like components.
B/T for EA3 and EA4 is 0.56 and 0.62, respectively. These values
are larger than the typical B/T for Sa galaxies (0.45) and
comparable to the median for S0's (0.63; Kent 1985). 
Despite the complications of fitting EA5, 
various modes of fitting the galaxy produce B/T $\sim 0.7$. Unless the bulge
and any underlying disk-like component
fade at dramatically different rates, which is unlikely 
given the relatively weak large-scale population gradients in these galaxies \citep{norton}, 
the descendants of these galaxies must be early type (S0 or E, if the disk-like
material is tidal debris that disperses or collapses onto the central component).

Given the asymmetric features, how reliable are these fits?
There are several ways to check the results for possible systematic errors.
First, we compare the fitted
analytical profiles to the radial surface brightness profiles obtained
from the isophote fitting procedure.
In Figure \ref{fig:profile}, we plot the radial surface brightness profiles
of a chosen model for each galaxy: $r^{1/4}$ profile for EA2 and EA5,
$r^{1/4}$ + exponential disk profile for EA3 and EA4, and 
the profiles obtained from the isophote fitting.
The differences between the data (ELLIPSE) and 
models (GALFIT) range mostly between $\pm 0.5$ mag/arcsec$^2$,
are not global, and reflect local asymmetric components.
%%%
Second, we examine the residual images obtained by subtracting 
the smooth and symmetric  models from the data (see Figure \ref{fig:residual}).
In all cases we see evidence for components beyond the bulge + disk model.
We then calculate how much light remains in the residual images
to quantify the goodness of the fit. 
The relative asymmetric light --- excess (deficit) --- within a 10$''$ radius is 
16(8)\%, 6(5)\%, 8(9)\%, 9(8)\% of the symmetric model components
for EA2-5, respectively. 
Most (50\% to 80\%) of the under(over)subtracted light comes from 
the central region within 0.5$''$, where the even a small amount of fractional 
deviation from the data can dominate the residual flux over the outer faint parts.
Except for EA2, the global under(over)subtractions are roughly the same
and localized fluctuations dominate the residuals, so we conclude that 
our global fits are reliable.
In the case with the most residual light (EA2),
24\% of the light cannot be explained by a symmetric model and the positive 
residuals dominate all over the galaxy.
Therefore, we cannot exclude the possibility that we are looking right along 
the interaction plane (the tidal debris are quite linear).
The point sources near the E+A bulges in the residual images
are discussed in \S\ref{sec:compact}.

\subsection{Morphologies: Asymmetric Components}
\label{sec:asymmetric}
So far we have fit symmetric smooth models with moderate success, but 
have found that asymmetric features are quite common in our sample.
Asymmetry, in particular lopsidedness, has been used to measure
disturbances in local ``normal'' disk galaxies \citep{RZ, ZR} and
correlates with recent star formation \citep{ZR, greg}. 
There are multiple ways in which one can quantify asymmetry, but here
we choose to follow what was done for local spirals by \cite{RZ}.
This measurement is based on the azimuthal Fourier decomposition of
the surface brightness along elliptical isophotes.

For the two most disk-like of the E+As (EA3 and 4), 
we calculate the Fourier decomposition of the F702W band surface
brightness distribution. 
We use a grid with 24 azimuthal and 36 radial bins from semi-major
axes of 4 to 200 pixels.
The center of the azimuthal grid is identified as the brightest central
point in the galaxy image.
Figure \ref{fig:fourier} shows the amplitudes of the various
first Fourier terms as a function of radius. In field spirals, $A_1 > 0.2$
is identified as strong lopsidedness, found in $\sim 20$\% of
the cases,  and interpreted as the result of a recent
interaction (other explanations, such as halo-induced disk sloshing 
have since been suggested; e.g., Levine \& Sparke 1998; Kornreich et al 2002).
EA3 has $A_1 \ll 0.2$  at all radii except in the transition region 
between the inner spiral arms and the tidal tails at $\sim 3-6$ kpc
and at large radii where the uncertainties are large.
Although EA4 has large $A_1$ within 3 kpc, which is
consistent with the appearance of the residual image (Figure \ref{fig:residual}), 
$A_1 < 0.2 $ for $1.5 < r_d < 2.5$, the range used in the study of 
field spirals. 
The lopsidedness of EA3 and EA4 is consistent with that of normal spiral 
galaxies despite the cataclysmic event that occurred $\sim$ Gyr ago 
indicated by the spectra. 
This result has two possible interpretations:  either these galaxies have
had sufficient time to relax and ``smooth out'' interaction-induced
asymmetries (e.g., Mihos 1995), or they are the result of encounters not
strong enough to cause major dynamical damage. Of course, the latter
possibility runs into the problem of how to trigger such a
massive starburst without dynamically disturbing the galaxy.

\subsection{Color Gradients}
\label{sec:color}
The color gradients of E+As are constraints on merger models and clues
as to what these galaxies will ultimately become. In Figure
\ref{fig:color}, we show the F702W$-$F439W color profiles  of
EA2-5, which are obtained by using the results from the ELLIPSE task and include
the A-type K-correction. Because of the
shallow exposure in F439W band, the color profiles are limited to 
$r \lesssim 2 - 3$ kpc, which is only 25\% of the radial coverage available in the 
red. To derive the
color profiles within the most central region ($<0.5''$), we use
the deconvolved images. However, because deconvolution
can produce large artificial fluctuations in low signal-to-noise data,  
we cannot use the deconvolved F439W profiles in the
outer regions of the galaxy. We compromise by using the deconvolved
images for $r < 0.5''$ and the
non-deconvolved images for $r > 0.5''$.

The overall colors of E+As are relatively blue globally due to the recent star
formation (bottom panel of Figure \ref{fig:color}). 
The radial extent of the blue colors confirms previous observations
\citep{Franx, Caldwell, norton}
that the recent star formation region is not confined within the innermost regions.
However, the color gradients,
especially within 1 kpc of the centers, are as diverse as the overall
morphologies.  
While, EA3 and EA5 have blue nuclei and become redder going outward, 
EA2 becomes bluer with radius, and EA4 shows a relatively flat profile.

The colors are the result of the complicated interplay between age,
metallicity and dust. The lack of HI in these systems \citep{chang,miller}
and of any patchiness in the images of EA2-5
argue against high levels of dust (but it is still possible that high density 
pockets of dust are present, particularly toward the nucleus of some of these systems).
With the exception of EA1, none of the E+As show the irregular, filamentary
structures expected from strong dust lanes. We thus conclude that the variety
of color gradients within the inner few hundred parsecs 
reflects variations of the spatial distribution of the young population, which
in some systems appears to be preferentially located near the center
of the galaxy and in others appears to avoid the center. 
Perhaps this reflects differences in the types of encounter involved
and its ability to drive true nuclear starbursts --- e.g., differences
between prograde and retrograde encounters \citep{Barnes_Hernquist},
major versus minor mergers \citep{Hernquist_Mihos95}, or differences in
the structural properties of the progenitor galaxies \citep{Mihos_Hernquist96}.

\subsection{Relationship to Fundamental Plane}
\label{sec:FP}
To investigate whether E+As can evolve into E/S0 galaxies, 
we compare the stellar kinematics and structural parameters
of E+As with ``normal'' early type galaxies. 
\citet{norton} found that the old component of E+A galaxies is offset 
(brighter by $\sim0.6$ mag) from the the local Faber-Jackson relation.
Using the structural parameters that can only be measured
using \emph{HST} imaging, we extend this comparison to 
various projections of the Fundamental Plane (hereafter FP) in Figure
\ref{fig:FP}. 
To compare our results with the FP of \citet{Jor96}, 
we correct these observables to our adopted cosmology 
($H_0 = 70 \mathrm{ \ km \ s^{-1}\ Mpc^{-1}}$, 
$(\Omega_M,\Omega_{\Lambda}) =(0.3, 0.7)$).
Changes in the cosmological parameters only affect zero points 
in the FP equation.
We use velocity dispersions from \cite{norton} for the K-star component 
and the structural parameters from a single $r^{1/4}$ model
\footnote{There is an issue as to which structural parameters 
(for the entire galaxy or the bulge only) one should use to construct 
the FP for S0 or disk galaxies.
However, we opt to use the ($r_e$, $\mu_e$) from a single $r^{1/4}$ profile, 
because we are comparing with \citet{Jor96}, who also used 
single de Vaucouleurs profiles for S0's.}.
We transform the F702W magnitude to a Gunn $r$ magnitude using
the average (F702W $-$ Gunn $r$) color for galaxies of various Hubble type 
\citep{fuku}.
The average ($r-\mathrm{F702W}$) colors range from 0.56 for elliptical 
to 0.51 for Scd  galaxies. Even for the extreme case of irregular (Im) 
galaxies, ($r-\mathrm{F702W}$) is different by only $\sim$ 0.1 magnitude 
from the average value of 0.54.

We show various projections of  the FP in Figure \ref{fig:FP}.
Figure \ref{fig:FP}(a) shows the face-on view of the FP given by 
$x = (2.21~log~r_e - 0.82~log\lan I \ran_e + 1.24~log~\sigma)/2.66$,
$y = (1.24~log \lan I \ran_e + 0.82~log~\sigma)/1.49$ (see \citet{Jor96}).
The dashed line indicates the bound set by the limiting magnitude, 
but the upper dotted boundary is not caused by a selection effect.
Figure \ref{fig:FP}(b) shows the edge-on view of the FP along the long
axis of the distribution, given by 
$y=1.24~log~\sigma - 0.82~log\lan I \ran_e$. 
Figure \ref{fig:FP}(c) shows the Faber-Jackson relation.
Figure \ref{fig:FP}(d) shows the correlations between $r_e$ and 
$\lan \mu_e \ran$.

The four E+As stand apart from the E/S0s in the edge-on view
of the FP, but otherwise populate the same general region of the 3-D volume. 
The most striking deviation of the E+As among the scaling laws lies 
in the $\mu_e-r_e$ correlation.
EA2-5 are more than a half magnitude brighter than the median E/S0 galaxies 
with the same effective radii. 
Especially, EA3 and EA4 have a large excess surface brightness 
over the dotted boundary of E/S0 galaxies in the FP. 
We measure the excess brightness to be 0.86 and 0.54 mag 
relative to the dotted upper boundary of the $\mu_e-r_e$ projection 
for EA3 and 4, respectively.
As seen Figure \ref{fig:FP}(c), although EA2-5 have intermediate
luminosities ($-22<M_r<-20$) in comparison to E and S0 galaxies,
they have a large mean surface brightness within the effective
radius. 

Will E+As fade onto the locus of E/S0's after a few Gyr?
We estimate the amount of fading 
within an effective radius using STARBURST99 models \citep{starburst99}.
Assuming solar metallicity, a Salpeter IMF, an underlying 
5 Gyr-old population (not evolving), and a single instantaneous starburst
with a mass fraction of 50\%, the E+As will fade $\sim 0.5$ magnitudes
during the first Gyr after the burst. 
This degree of fading places EA2, EA4, and EA5 on the E/S0 locus, but not EA3.
This model is oversimplified, however. 
The evolutionary tracks in the $\mu_e-r_e$ projection will proceed from
upper-left to lower-right, because of the decline of the mean surface brightness 
within $r_e$ both due to fading and the inclusion of the larger fraction of 
the galaxy.
But, calculating these effects requires a detailed knowledge
of the spatial distribution of the young and old populations.

\subsection{Comparison with to ``Core'' and ``Power-law'' Ellipticals}
\label{sec:nuker}
Does the consistency in global surface brightness profiles between E+As
and normal early-type galaxies (as discussed in \S \ref{sec:FP})
extend to the innermost radii? The recently discovered dichotomy
in the central surface brightness profiles (``core'' vs. ``power-law''; Lauer et al. 1995; 
Faber et al. 1997) of normal early-type provides a framework for
this comparison.
In Figure \ref{fig:nuker}, we plot the F702W surface brightness profiles for our E+As 
and the F555W profiles of \citet{f97} sample (assuming $V - R = 0.5$ and adjusting for
differences in adopted cosmology).
The apparent lack of any central ``break'' and associated transition to a shallow cusp
in the E+A profiles leads us to classify the E+As as ``power-law'' galaxies.
Not only are the profiles shapes consistent, but the total luminosities of these E+As 
lie within the range spanned by normal power-law early-types, and any fading will
place them even more securely in that range.

Although the profile shapes and total luminosities of E+As are 
consistent with those of power-law early-types, Figure \ref{fig:nuker}
shows that the E+A surface brightness are generally higher. 
EA2 is consistent with the highest surface brightness normal power-law galaxies,
and the other three E+As have even higher surface brightness. If fading 
of the young stellar population does not significantly alter the profile shape,
then these galaxies will retain their power-law profiles.
While we cannot be certain that the profiles will remain unchanged, 
the current inner profiles and total luminosities are consistent with
E+A evolution into normal power-law early-type galaxies.

\subsection{Discovery of Young Star Clusters}
\label{sec:compact}
A specific type of asymmetric small-scale feature is the compact 
sources most visible in the residual images.
We identify a set of these sources using SExtractor \citep{bertin} 
to analyze both the combined and residual images. 
Because we expect stellar clusters to be unresolved at these distances 
(one pixel corresponds to $\sim 65-95$ pc at $z = 0.07 - 0.12$), 
confusion between real sources 
and hot pixels or background fluctuations is a serious problem.
After testing various detection criteria and comparing visually, 
we decided to consider an object real only if it has at least 
three adjacent pixels that are each at a flux level 3 times above 
the local background rms. 
To estimate the local background level, we use a background mesh size
for SExtractor of $\sim 4 - 8$ pixels, depending on the image. 
The residual images are used to detect compact sources close to the galaxy center, 
where the rapidly varying galaxy light complicates detections 
in the original images. 

Using these criteria, we detect 35, 9, 29, 10 and 1 point-like sources 
within the PC images of EA1-5, respectively.
These objects may be foreground stars, background compact sources, 
giant HII regions,  or star clusters.
Unfortunately, due to the insensitivity of the F439W band,
we are able to obtain colors for only a few of the objects.
The lack of  emission lines in the spectra of  these E+A galaxies 
\citep{zab96,norton} 
suggests that these  galaxies are not littered with giant HII regions.
The spatial distribution of the compact objects suggests that 
the majority of these sources are associated with the E+As.
For example, in EA1 almost all of the point sources are located around
EA1 and its companion, and three of them lie in the bridge-like
structure connecting the two galaxies. 
In EA2, some objects lie in the elongated tidal tails. 
In EA3, five to six unresolved objects
surround the nucleus of EA3 and form a concentric circle.
We apply a Kolmogorov-Smirnoff (K-S) one-sample test 
to find the probability that the radial distributions of the compact 
sources are drawn from a random distribution. 
We show the cumulative distribution of the angular distances from the center 
of each galaxy in Figure \ref{fig:kstest}. 
There is almost zero probability that the sources follow
the random distribution, indicating that the detected compact sources 
are truly associated with the E+A galaxies.
Because of the post-starburst nature of E+A galaxies and the
discovery of numerous young clusters in ongoing mergers
(e.g., Holtzman et al. 1992; Whitmore et al. 1993), it is likely that 
the majority of these objects are relatively young star clusters.

Like the morphological diversity of the E+As themselves, the number of cluster candidates
varies widely among EA1-5. In particular, 
the difference between EA3 and EA5 is remarkable because these
otherwise appear to have comparable masses (the velocity dispersions of
their old populations are the same and the luminosity ratio is $\sim 0.5$). 
The principal difference between these systems is that EA5 is more distant. 
To determine whether the distance difference 
is responsible for the difference in detected sources,
we model the detectability of the point sources as a function of distance.
For a given size (half-light radius) and brightness, we
generate two dimensional surface brightness distributions of the
model clusters for a King profile \citep{king} with concentration $c = 0.5 - 2.5$.
We convolve the model profiles with the PC PSF generated from
the TinyTim software \citep{Krist99}. 
All models and PSFs are constructed at a resolution of 1/10th of a 
PC pixel, re-binned to the original image size, and smeared using the pixel
smearing kernel.
After adding the Poisson noise and sky background, we attempt to detect
the cluster using the same detection criteria that we apply to the original images.
The position on the PC and the spectral variation of the
PSF do not affect our detection threshold.

The number of detected pixels above 3$\sigma$ is nearly insensitive 
to the size of the sources and the shape of the profile in the subpixel regime.
In other words, the images of point sources give us little or no information
about their spatial structures, but they can be detected if they are sufficiently bright.
We adopt the detection limit as the magnitude at which the number of detected pixels
above 3$\sigma$ drops below three pixels.
The number of compact sources above EA5's detection limit ($M_R<-12.5\pm0.25$)
and well outside the E+A are 6 to 16 for EA3 and is 1 for EA5.
The uncertainty in the detection limit arises from uncertainties 
in the aperture correction and photometry.
Assuming EA5 has the same number of the compact sources as EA3,
we would expect to detect 3 to 8 sources in EA5 after rough scaling 
of the projected area occupied by the detections around each galaxy.
The statistics of our sample thus preclude us from distinguishing 
between the cluster populations of EA3 and EA5.
If the difference is significant, it might reflect a difference in the starburst  
(strength or progenitors) or an age difference.
The latter possibility is consistent with the observation
that EA5 has the reddest $D_{4000}$ and is the most symmetric among our sample. 

Matching the observed colors of the cluster candidates 
with the predictions of stellar population synthesis models can
help break the degeneracy between the age and starburst strength 
of our post-merger E+As. 
We show the evolution of the color and brightness of a 
simple stellar population based on Starburst99 models 
\citep{starburst99} in Figure \ref{fig:evolution}.
Because of the unknown SEDs of our sources, the K correction of the
observed colors is problematic.
To avoid applying an uncertain K correction, 
we calculate the redshifted (F439W-F702W) colors from the model
spectra provided from Starburst99.
Superposed on the various models, we show the range of the colors observed 
and the corresponding range of ages for each of the five point-like sources 
in EA1 with $\sigma(\mathrm{F439W-F702W}) < 0.25$ and sufficient S/N.
They are identified 
by eye in the F439W band and are not located in the crowded region.
Because the absence of the O, B stellar signatures \citep{zab96,norton} suggest that the ages of the 
clusters are older than $10^7$ yr,
we rule out the solutions in the double trough of the model.
Assuming a Salpeter IMF that extends from 1 to 100 $M_\odot$, 
solar metallicity ($Z=0.020$), and an instantaneous
starburst, we estimate the ages of the candidate clusters to be between $\sim$ 30 Myr
and 500 Myr.

We applied the following statistical test to constrain the time since the starburst
under the assumption that all the candidate clusters formed 
at the same time and have a spread of estimated age due to
photometric errors.
For each model age, we draw a thousand samples of five clusters
each with colors scattered according to a Gaussian of dispersion 0.2 mag.
We calculate the percentage of simulations that have a median color 
as far from the true color as the observed median color. 
If that number is 5\% or less, the model is rejected. 
Because of the slow evolution of colors in the
relevant age range, we can constrain the age of clusters
(time elapsed since the starburst) to only $\sim$ 35 $-$ 450 Myr at the 95\%
confidence level (see the horizontal line in Figure \ref{fig:evolution}).
While our constraint on the age is drawn from only a few cluster candidates,
this range is as good as the typical age ranges that \citet{Leonardi_Rose}
derived using CaII and $H\delta/\lambda4045$ indices from the integrated
spectra of E+A galaxies.
%%%
Because any reddening by the host galaxy will make the sources look older, 
our estimated ages are upper limits on the true ages.
However, it is unlikely that the reddening affects our results
seriously because the five sources are at large radii, where the extinction
should be minor compared to the galaxy center.
To improve upon the age estimate, in particular to reduce contamination,
to correct the reddening, and to get more precise ages 
for the sources, ideally we would need deep multicolor photometry to produce
a two-color diagram like Fig. 6 in \citet{jharris}.  
However, we can see from the present analysis that we would benefit from
just more clusters with colors.
We conclude that the derived ages are consistent with 
what we expect on the basis of the integrated galaxy spectra, 
again supporting the claim that these are stellar clusters
associated with the E+A phenomenon, and suggesting that this is part of 
a population of clusters formed during the E+A phase.

We compare the R band cluster luminosities of EA1, EA3,
the Milky Way globular clusters \citep{harris} and the clusters 
in starbursting, merging systems (e.g., NGC3597, Carlson et al 1999).
If the compact sources in the E+As are clusters, they are 
much brighter ($M_R \sim -13$) than Galactic globular clusters ($M_R > -11$) and 
similar to clusters in galaxies with on-going starburst.
The latter agreement supports the interpretation that we have identified 
the bright end of a population (e.g., $M_R \sim -14$ in NGC3597) of star clusters 
formed during a starburst that occurred $< 1$ Gyr ago. 
A fading of several magnitudes is required for these systems to resemble 
the massive end of the Milky Way cluster population.

\subsection{Implications for High Redshift Interacting Galaxies}
\label{sec:highz}
The difficulty we have experienced in determining whether these E+As
have a disk component, have tidal components, and are asymmetric suggests
that at high redshift these galaxies might be classified as ``normal''
morphologically. 
For example, the tidal tails that connect onto what may be a disk would 
naturally be interpreted as spiral arms if one is not able to trace
the tails out to large radii.
To better understand this effect, we rebin the F702W band image of EA3 
and fade it artificially according to our adopted cosmology
to mimic its appearance at higher redshift (Figure \ref{fig:highz}).
The rebined images are convolved with the PSF, and 
the sky background and noise proportional to the exposure time are
added to the redshifted images.
For each image, we assume the same exposure time (2100s) and average sky brightness. 
The tidal features are lost at a
redshift over $\sim 0.5$ (see also Mihos 1995; Hibbard \& Vacca 1997), and EA3 appears to be a normal spiral or S0 galaxy.
The mean surface brightness of the
tidal tails of EA3 is $\sim 24 - 25$ $\mathrm{mag~arcsec^{-2}}$.
This exercise suggests that some of the disky E+A galaxies found
frequently in distant clusters 
might be misclassified as non-interacting disk galaxies and nevertheless have
tidal features like EA2-4. The lack of apparent tidal features and of asymmetry at high redshift
should not be interpreted directly as an absence of interactions.

\section{Summary}
\label{summary}

Using \emph{HST} imaging, we have obtained high resolution images of
five E+Ai galaxies.
Our results, when coupled with studies of E+A kinematics, star formation 
histories, and gas content, argue that E+As are in transition 
between late and early types, a missing link in the evolution of galaxies.
Our principal findings are:

% disturbed morphology
1. Four (EA1-4) of the five E+As are morphologically disturbed (with tidal tails and
shell-like structures), consistent with being relics of
galaxy-galaxy interactions.
The dramatic tidal features found in EA1, 2, and 3 further confirm that
galaxy-galaxy mergers are at least one mechanism that
triggers the starburst that leads to the  E+A phase.

% morphological diversity
2.  E+As have diverse morphologies. Even in this small sample, 
one highly disturbed galaxy cannot be morphologically classified
(EA1) and one regular galaxy could plausibly be classified as a barred
S0 galaxy (EA5). 
Two other galaxies (EA3 and 4) seem to contain disk components.
The striking aspect of this morphological variety is
that all of these galaxies have been selected on the basis of their very similar spectra.

% bulge dominated
3. The bulge fractions of these systems are consistent with elliptical and S0 galaxies.
E+As have a central structure consistent with ``normal'' power-law early type
galaxies \citep{f97}, although at present the E+As have significantly 
higher surface brightnesses.
This is the first direct confirmation that power-law elliptical galaxies
can be formed from gas-rich mergers.

% color profile : diversity and red core
4. E+As are blue due to recent star formation, 
but their color gradients in the central region are as diverse 
as their morphologies. Two galaxies (EA3 and 5) have extremely 
blue cores that redden toward the outer regions, one (EA4) has a relatively flat
profile, and one (EA2) has a redder core within a few kpc. The red core
is particularly puzzling from the standpoint of merger models that
predict gas inflow into the center and a subsequent
starburst. It is possible that the red color arises from dust extinction, 
but we see no evidence for the filamentary structure usually associated
with dust lanes.

% fundamental plane : endproduct?
5. E+As stand apart from the E/S0s in the Fundamental Plane
mainly due to their high central surface brightness. 
Fading of the young stellar population and a corresponding increase in $r_e$ 
(if the young population is more concentrated than the underlying population) 
will cause the systems to migrate toward the locus of E/S0s in time. 
Without a detailed knowledge of the distribution of young and old stars 
it is not possible to model this evolution precisely.

% point-like sources : star cluster?
6. We find compact sources associated with E+A galaxies. 
Although some of these sources 
could be foreground stars, giant HII regions, or background objects, we
conclude that the majority are star clusters that formed during 
the starburst phase,
such as those found in nearby starburst galaxies.
These systems are similar in luminosity to the
brightest clusters in NGC 3597 \citep{carlson} and have ages
consistent with that expected for the time elapsed since the starburst (several 10$^8$ yrs).
They are much brighter ($M_R \sim -13$) than 
Galactic globular clusters, but will fade over a Hubble time to be consistent
with the most massive Milky Way clusters. Further study of these systems
might provide the elusive chronometer for lifting the age/burst-strength
degeneracy for post-merger galaxies.

% cautionary remark on the higher redshift
One valuable lesson of our work is that despite the recent
interactions that at least some of these E+As
have experienced, one of the standard tests for interactions (the
lopsidedness measure based on the azimuthal Fourier
decomposition of the isophotes) fails to identify these systems as 
having experienced a recent interaction. The tidal features are sufficiently
faint and the inner surface brightness profiles sufficiently regular that
the galaxies are measured to be symmetric. Such indicators should
therefore be used with caution on large-scale surveys and at high-redshifts
because they obviously miss at least some recently interacting systems.
Furthermore, in a system that one does not suspect to be interacting,
the tidal tails can easily be misidentified as spiral arms.

% future work : ACS
This work described the observations of 
the five bluest E+A galaxies in LCRS sample with
\emph{HST}/WFPC2. To confirm the various results presented here and
determine whether this subset was biased by the color selection
requires a study of the remainder of the LCRS sample.
We are currently obtaining \emph{HST}/ACS imaging
of those galaxies.

\acknowledgements
We thank Chien Peng for help in using his program (GALFIT) and useful
discussions about galaxy morphology. We also thank Greg Rudnick for providing
the azimuthal Fourier transform routines.
YY and AIZ acknowledge support from NSF grant AST-0206084, NASA LTSA grant \#NAG5-11108, 
and HST grant GO-06835.01-95A.
J.C.M. acknowledges support by the NSF through CAREER grant
AST-9876143 and by a Research Corporation Cottrell Scholarship.
Support for proposal GO-6835 was provided by NASA through 
a grant from the Space Telescope Science Institute, which is operated 
by the Association of Universities for Research in Astronomy, Incorporated, 
under NASA contract NAS5-26555.

%%%%%%%%%%%%%%%%%%%%%%%%%%%%%%%%%%%%%%%%%%%%%%%%%%%%%%%%%%%%%%%%%%%%%%
%	bibliography
%%%%%%%%%%%%%%%%%%%%%%%%%%%%%%%%%%%%%%%%%%%%%%%%%%%%%%%%%%%%%%%%%%%%%%

\clearpage

%%%%%%%%%%%%%%%%%%%%%%%%%%%%%%%%%%%%%%%%%%%%%%%%%%%%%%%%%%%%%%%%%%%%%%
%	Table  
%%%%%%%%%%%%%%%%%%%%%%%%%%%%%%%%%%%%%%%%%%%%%%%%%%%%%%%%%%%%%%%%%%%%%%

%Table 1
\begin{deluxetable}{crrcccrrcrrc}
\tablewidth{0pt}
\tabletypesize{\scriptsize}
\tablecaption{Properties of galaxies\label{tab:basic}}
\tablehead{
\colhead{ID}		&\colhead{R.A.}		&
\colhead{Dec.}		&\colhead{$z$} 		&
\colhead{$kpc/''$}	&\colhead{$f_A$ \tablenotemark{a}}	&
\multicolumn{2}{c}{$\lan V_{rot} \ran$ \tablenotemark{a}}	&\colhead{}	&
\multicolumn{2}{c}{$\sigma~(\mathrm{kms^{-1}})$ \tablenotemark{a}} 	&
\colhead{cluster} \\
\cline{7-8} \cline{10-11}
\colhead{} 	&\colhead{(1950.0)}	&\colhead{(1950.0)} &
\colhead{}	&\colhead{}			&\colhead{} 		&
\colhead{Old}	&\colhead{Young}&\colhead{}			&
\colhead{Old}	&\colhead{Young}&
\colhead{?}
}
\startdata 
EA1&10 58 48.97&-11 54\B9.80&0.0746&1.417&0.80&24 \scr$ +57 \atop -54 $&19 \scr$ +17 \atop -17 $&&\phn35 \scr$ +39 \atop -35 $&\phn23\scr$ +53 \atop -23 $&N \\
EA2& 2 15 43.24&-44 46 36.70&0.0987&1.823&0.42&23 \scr$ +16 \atop -20 $&22 \scr$ +24 \atop -25 $&&   202 \scr$ +17 \atop -16 $&   193\scr$ +32 \atop -46 $&N \\
EA3&12\B6 31.34&-12\B5 55.40&0.0810&1.527&0.59&23 \scr$ +17 \atop -23 $&25 \scr$ +18 \atop -17 $&&   120 \scr$ +22 \atop -20 $&\phn56\scr$ +35 \atop -32 $&N \\
EA4& 3 58 23.42&-44 43 40.29&0.1012&1.864&0.56&1  \scr$ +\B8\atop-\B9 $&35 \scr$ +15 \atop -16 $&&   131 \scr$+\B9 \atop-\B9 $&   246\scr$ +24 \atop -25 $&Y \\
EA5& 1 56\B0.12&-44 51 49.00&0.1172&2.119&0.42&33 \scr$ +24 \atop -25 $&52 \scr$ +34 \atop -18 $&&   120 \scr$+\B7 \atop-\B8 $&\phn94\scr$ +34 \atop -41 $&N 
\enddata
\tablenotetext{a}{Data from \citet{norton}}
\end{deluxetable}

%Table 2
\begin{deluxetable}{ccccccccccccc}
\tablewidth{0pt}
\tabletypesize{\small}
\tablecaption{Magnitudes and colors of E+A galaxies\label{tab:mag}}
\tablehead{
\colhead{}&\multicolumn{2}{c}{F439W}&\colhead{}&\multicolumn{2}{c}{F702W}&\colhead{}&\colhead{F439W$-$F702W}\\
\cline{2-3} \cline{5-6} \cline{8-8}
\colhead{ID}&\colhead{$m ~(< 4.5'')$}&\colhead{$M$}&\colhead{}&\colhead{$m~(< 4.5'')$}&\colhead{$M$}&\colhead{}&\colhead{$(r<r_e)$}
}
\startdata
EA1&17.95$~\pm~$0.02&-19.97&&17.18$~\pm~$0.01&-20.40&&--- \\
EA2&18.09$~\pm~$0.02&-20.50&&16.44$~\pm~$0.01&-21.70&&1.10$~\pm~$0.01\\
EA3&16.75$~\pm~$0.01&-21.52&&15.34$~\pm~$0.01&-22.49&&0.80$~\pm~$0.01\\
EA4&17.47$~\pm~$0.01&-21.17&&15.86$~\pm~$0.01&-22.31&&1.18$~\pm~$0.01\\
EA5&18.58$~\pm~$0.04&-20.51&&16.83$~\pm~$0.01&-21.67&&0.99$~\pm~$0.01
\enddata
\end{deluxetable}

%Table 3
\begin{deluxetable}{ccccccccc}
\tablewidth{0pt}
\tabletypesize{\small}
\tablecaption{K corrections for A dwarf and K giant\label{tab:kcorr}}
\tablehead{
\multicolumn{2}{c}{} & \multicolumn{2}{c}{A dwarf }& 
\multicolumn{1}{c}{} & \multicolumn{2}{c}{K giant } \\
\cline{3-4} \cline{6-7}
\colhead{ID}&	\colhead{$z$}&	\colhead{$K$(F439W)}&	\colhead{$K$(F702W)}&
					\colhead{}&		\colhead{$K$(F439W)}&	\colhead{$K$(F702W)} 
}
\startdata
EA1&0.0746&0.136~~$\pm$0.035&-0.148~~$\pm$0.022&&0.755~~$\pm$0.123&0.042~~$\pm$0.032\\
EA2&0.0987&0.233~~$\pm$0.038&-0.192~~$\pm$0.028&&0.991~~$\pm$0.168&0.065~~$\pm$0.038\\
EA3&0.0810&0.157~~$\pm$0.036&-0.160~~$\pm$0.024&&0.823~~$\pm$0.136&0.048~~$\pm$0.034\\
EA4&0.1012&0.246~~$\pm$0.039&-0.196~~$\pm$0.029&&1.016~~$\pm$0.172&0.068~~$\pm$0.039\\
EA5&0.1172&0.340~~$\pm$0.042&-0.225~~$\pm$0.033&&1.176~~$\pm$0.192&0.084~~$\pm$0.043

\enddata
\end{deluxetable}

\newpage
\newpage
%Table 4
\begin{deluxetable}{ccccccccc}
\tablewidth{0pt}
\tabletypesize{\small}
\tablecaption{Structural parameters\label{tab:param}}
\tablehead{
\colhead{}			&\multicolumn{3}{c}{$r^{1/4}$-Law} &
\colhead{}			&\multicolumn{4}{c}{S\'ersic $r^{1/n}$-Law} \\
\cline{2-4} \cline{6-9} 
\colhead{ID}		&\colhead{$r_e ~(kpc)$}			&
\colhead{$\mu_e$}	&\colhead{$\chi^2_{\nu}$}	&
\colhead{}			&
\colhead{$r_e ~(kpc)$}		&\colhead{$\mu_e$}	&
\colhead{$n$}		&\colhead{$\chi^2_{\nu}$	}
}
\startdata 
EA2&2.74$~\pm~$0.14&19.99$~\pm~$0.09&1.72&&4.49$~\pm~$0.98&21.10$~\pm~$0.39&6.22$~\pm~$0.53&1.62\\
EA3&1.79$~\pm~$0.07&18.56$~\pm~$0.07&2.08&&3.74$~\pm~$0.64&20.19$~\pm~$0.28&8.73$~\pm~$0.45&1.51\\
EA4&2.07$~\pm~$0.07&19.22$~\pm~$0.06&1.61&&2.43$~\pm~$0.24&19.61$~\pm~$0.17&5.10$~\pm~$0.25&1.56\\
EA5&1.62$~\pm~$0.06&18.85$~\pm~$0.07&1.49&&2.39$~\pm~$0.38&19.70$~\pm~$0.31&6.40$~\pm~$0.56&1.45
\enddata
\end{deluxetable}

%Table 5
\begin{deluxetable}{ccccccccc}
\tablewidth{0pt}
\tabletypesize{\small}
\tablecaption{Bulge-disk decompositions\label{tab:decomp}}
\tablehead{
\colhead{}			&\multicolumn{2}{c}{Bulge} &
\colhead{}			&\multicolumn{2}{c}{Disk} \\
\cline{2-3} \cline{5-6} 
\colhead{ID}		&\colhead{$r_e ~(kpc)$}			&
\colhead{$\mu_e$}	&\colhead{} 				&
\colhead{$r_d ~(kpc)$}		&\colhead{$\mu_d$}	&
\colhead{$\chi^2_{\nu}$}	&\colhead{B/T}
}
\startdata 
EA3&0.69$~\pm~$0.03&17.02$~\pm~$0.05&&3.57$~\pm~$0.22&19.30$~\pm~$0.14&1.58&0.56\\
EA4&1.11$~\pm~$0.05&18.11$~\pm~$0.05&&3.44$~\pm~$0.16&19.50$~\pm~$0.12&1.48&0.62\\
EA5&1.14$~\pm~$0.03&18.23$~\pm~$0.04&&4.01$~\pm~$0.45&21.00$~\pm~$0.21&1.45&0.70
\enddata
\end{deluxetable}

%%%%%%%%%%%%%%%%%%%%%%%%%%%%%%%%%%%%%%%%%%%%%%%%%%%%%%%%%%%%%%%%%
% FIGURE
%%%%%%%%%%%%%%%%%%%%%%%%%%%%%%%%%%%%%%%%%%%%%%%%%%%%%%%%%%%%%%%%%

\begin{figure}
\figurenum{1}
\plotone{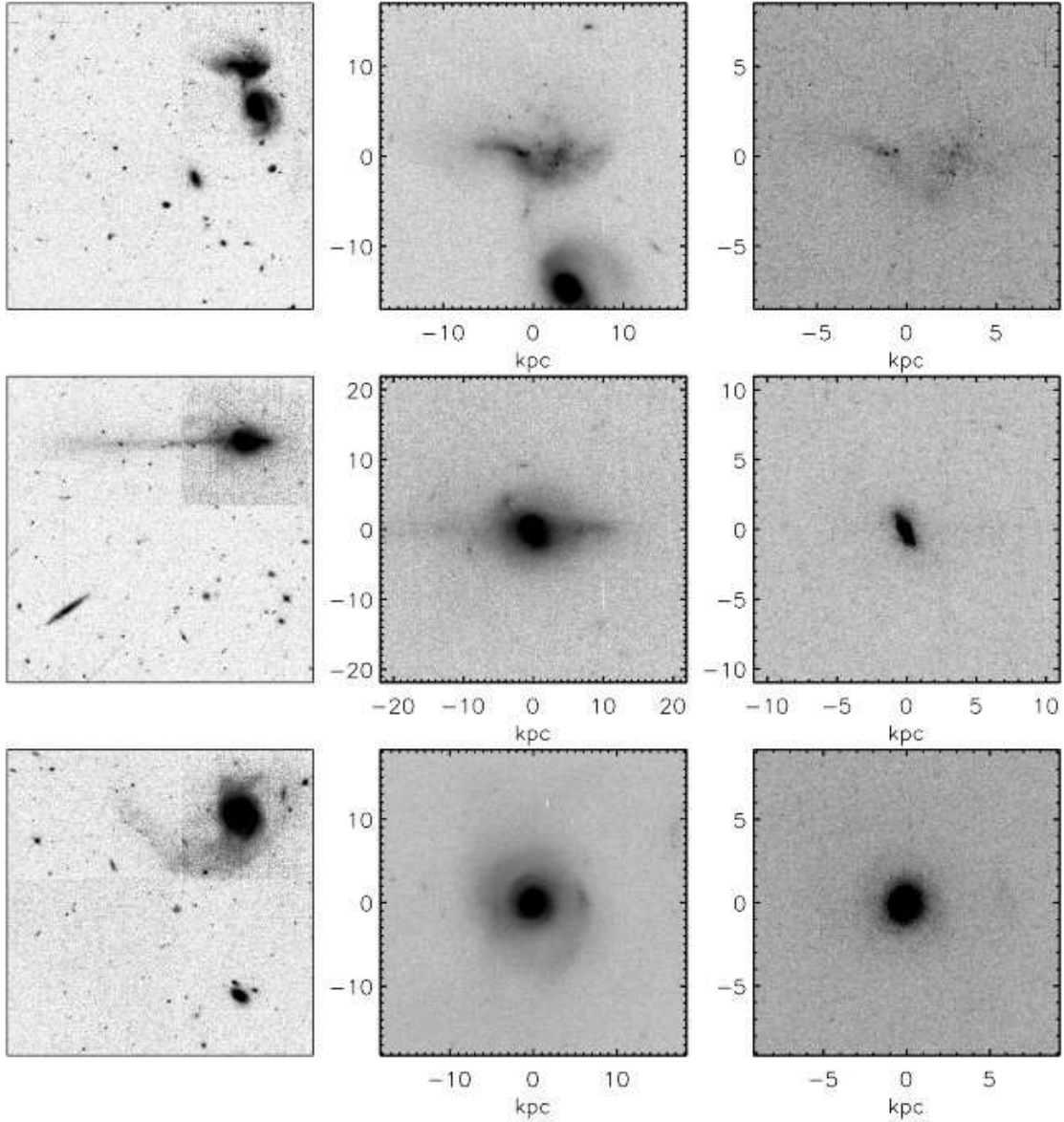} 
\caption{\emph{HST} WFPC2 images of five E+A galaxies. From top to bottom, EA1
  to EA5. (\emph{left column}) WFPC2 mosaic images in high
  contrast. Notice the dramatic tidal tails. (\emph{middle column})
  High resolution F702W band PC images in low contrast.
  (\emph{right column}) Central regions in F439W band.
  The size of each field is $\sim80''$, $24''$, and $12''$ (left to right), 
  and the pixel scale in the middle and right columns is $0''.046$/pixel.}
\label{fig:tile}
\end{figure}

\begin{figure}
\figurenum{1}
\plotone{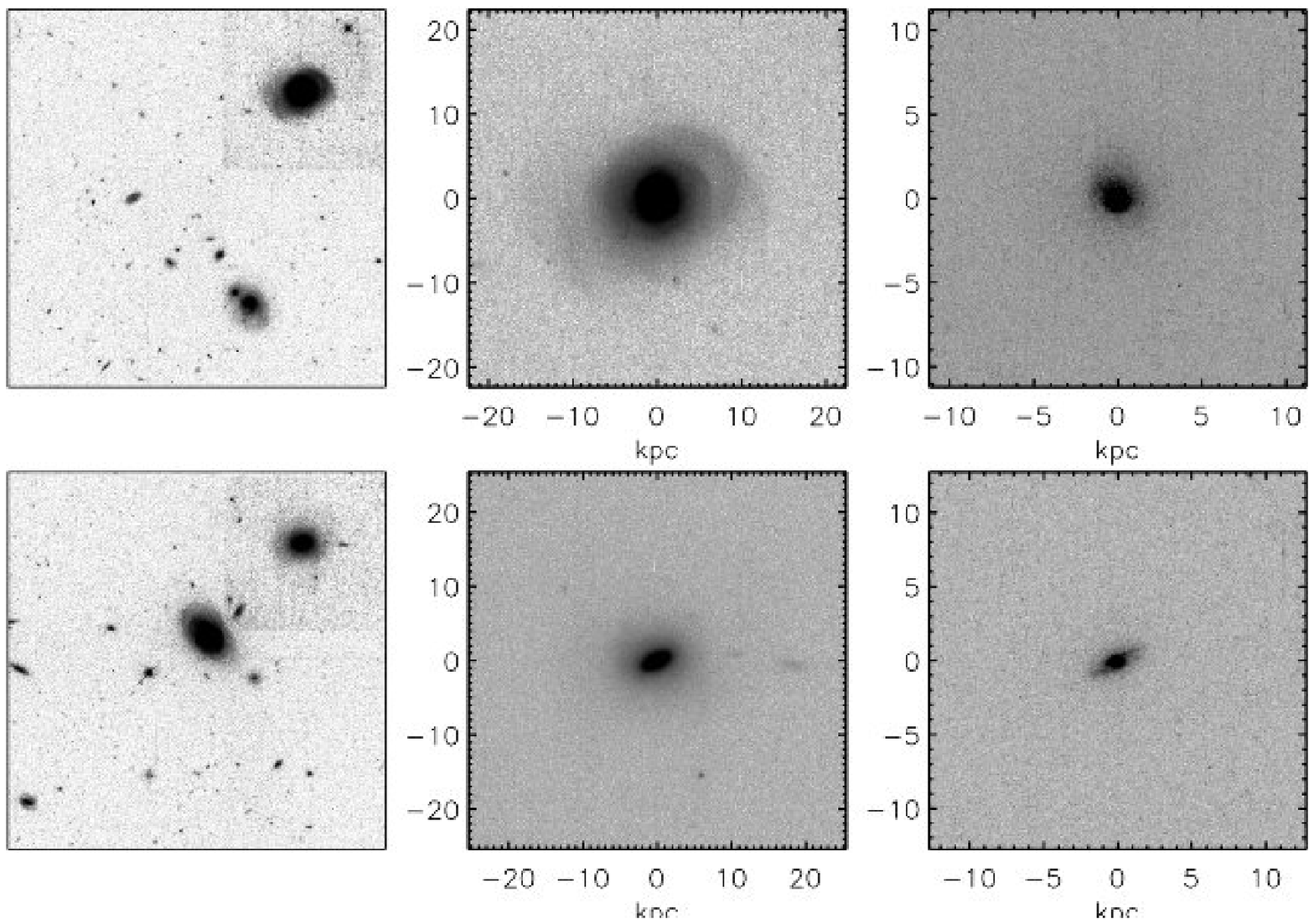}
\caption{Continued.}
\end{figure}

\begin{figure}
\figurenum{2}
\plotone{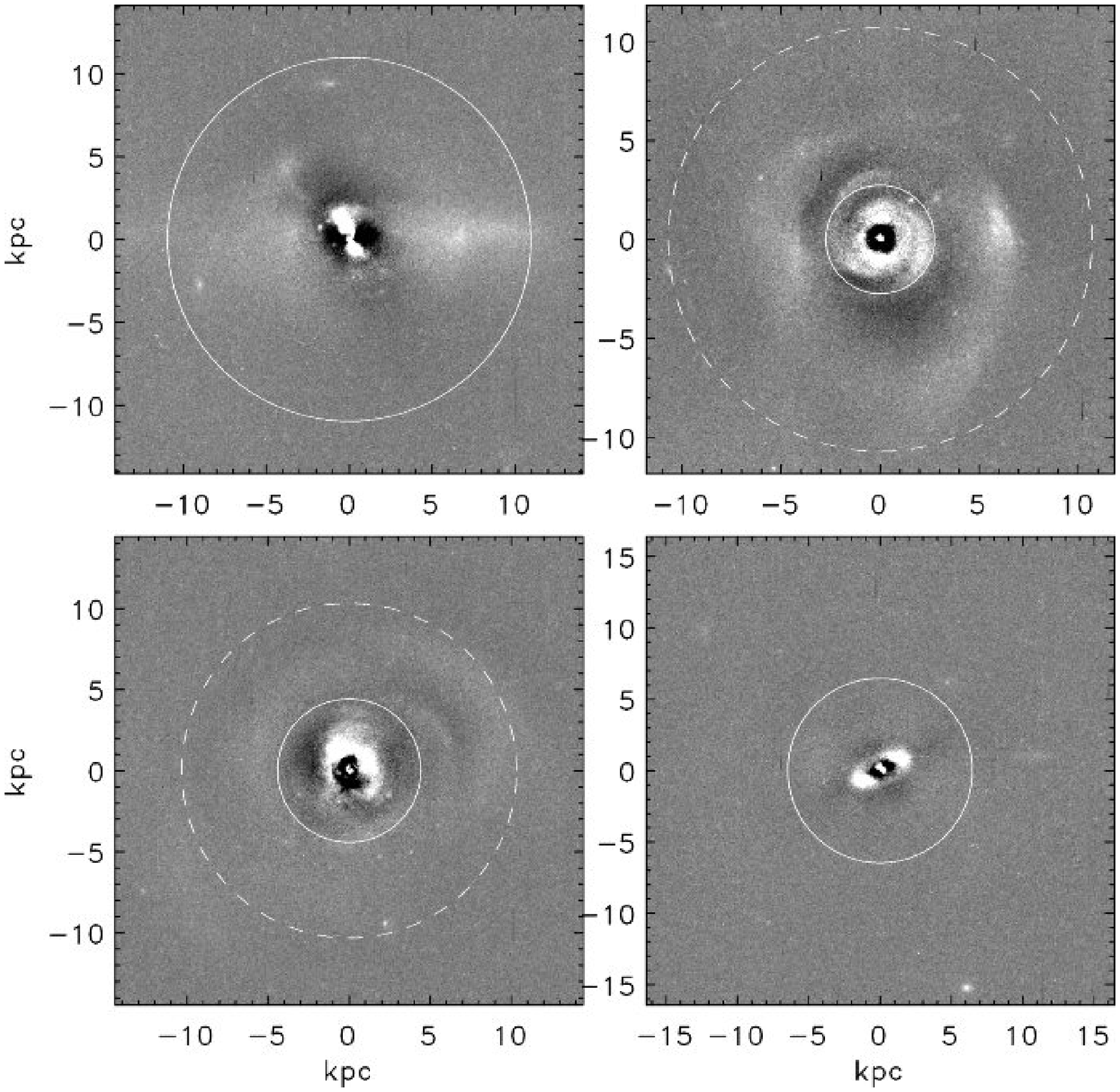}
\caption{Residual images ($15.5''\times15.5''$) obtained by subtracting
  the smooth and symmetric model images from the data. 
  EA3 and EA4 : $r^{1/4}$ bulge + exponential disk model.
  EA2 and EA5 : $r^{1/4}$ bulge only.
  The solid and dashed circles represent $4 r_e$ 
  and $3 r_d$ (if bulge+disk decomposition was done), respectively.
  The relative asymmetric light --- excess (deficit) --- within a 10$''$ 
  radius is 16(8)\%, 6(5)\%, 8(9)\% and 9(8)\% of the symmetric model 
  components for EA2-5, respectively.
\label{fig:residual}}
\end{figure}

\begin{figure}
\figurenum{3}
\epsscale{0.9}
\plotone{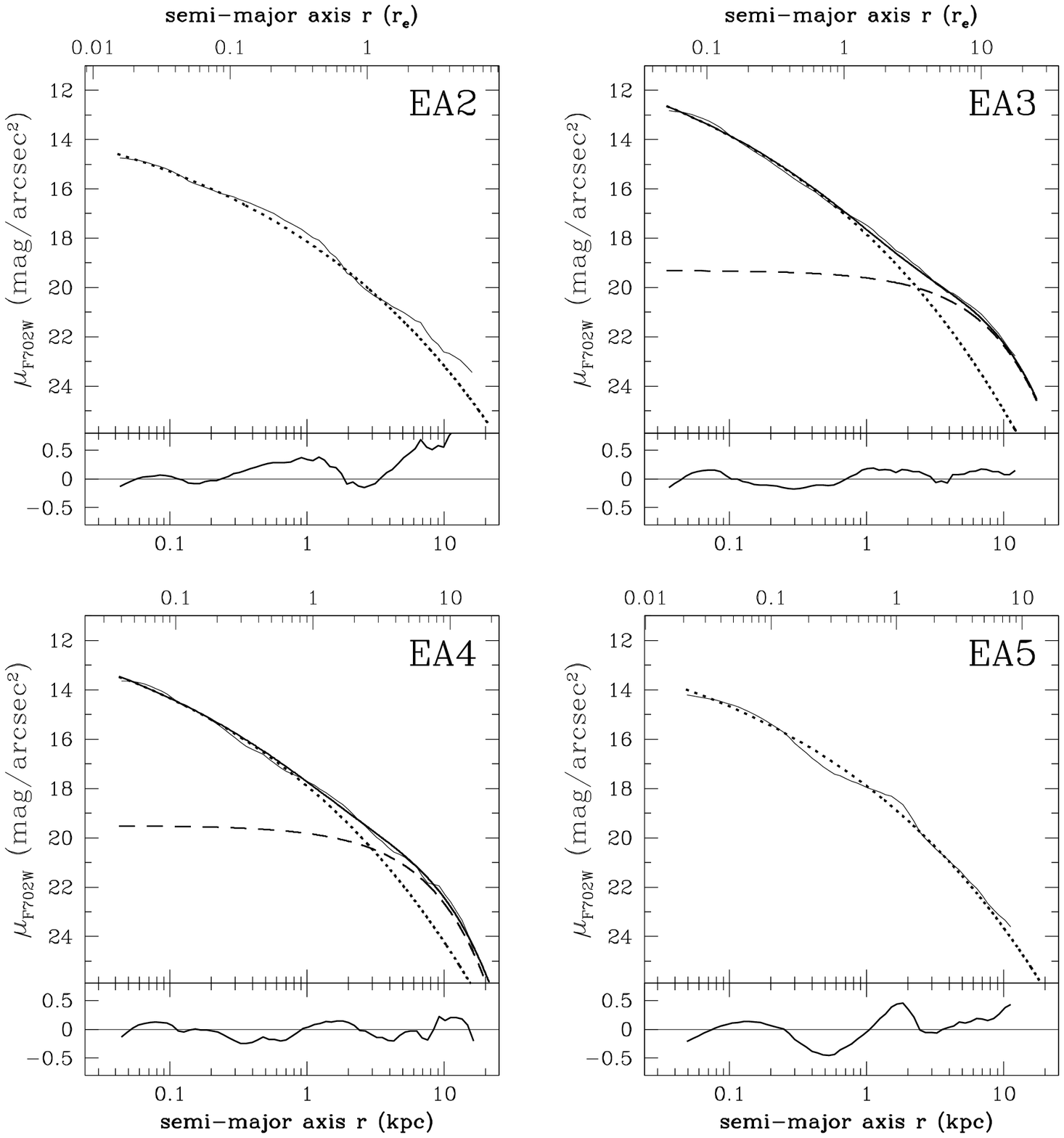}
\caption{F702W radial surface brightness profiles. 
  The thin solid lines show the radial profiles obtained from ELLIPSE,
  the dotted lines show the bulge components fitted to 
  the $r^{1/4}$ law, and the dashed lines show the disk/tidal components
  fitted to the exponential law. 
  The thick solid lines are the superposition of the bulge and disk components.
  Pure bulge + disk decomposition was done only for EA3 and EA4
  (fitting EA5 was more complicated, see text and Figure \ref{fig:ea5}).
  The top axis of each panel is the effective radius $r_e$ 
  of the bulge component of each galaxy.
  Bottom panels show the difference between the data (ELLIPSE) and fits (GALFIT).}
\label{fig:profile}
\end{figure}

\begin{figure}
\figurenum{4}
\plotone{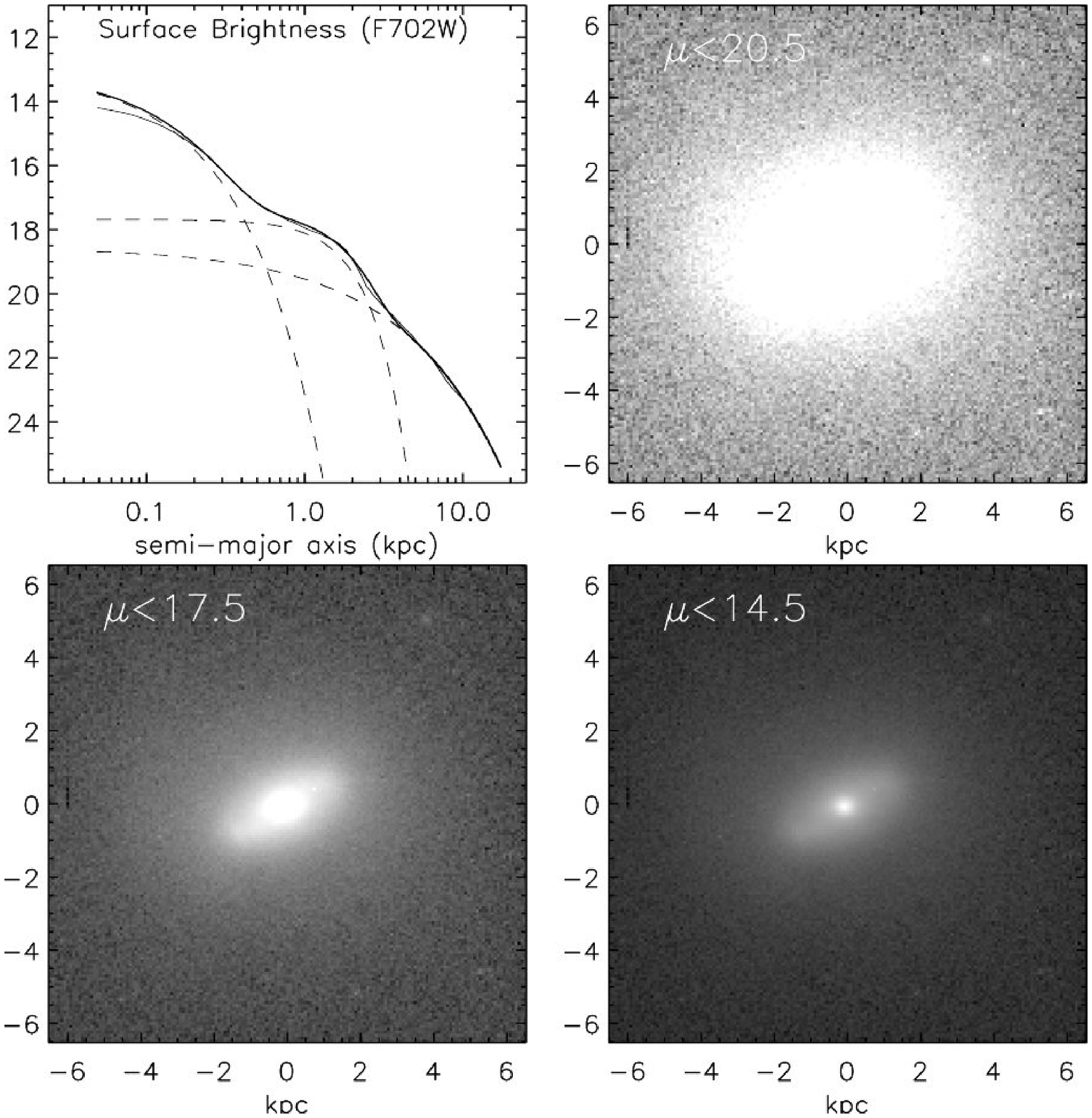}
\caption{EA5 galaxy in different contrasts. EA5 shows three components ---
  an extended light distribution in the outer part, a bar structure, 
  and a very bright blue central nucleus. (\emph{Upper left panel}) 
  Three-component decomposition of EA5. The dashed lines represent each 
  component. Solid thin and thick lines show the data and the sum of the
  three model components ($n = 1.1, ~0.5$, and $1.5$ from inside to outside),
  respectively.}
\label{fig:ea5}
\end{figure}

\begin{figure}
\figurenum{5}
\plotone{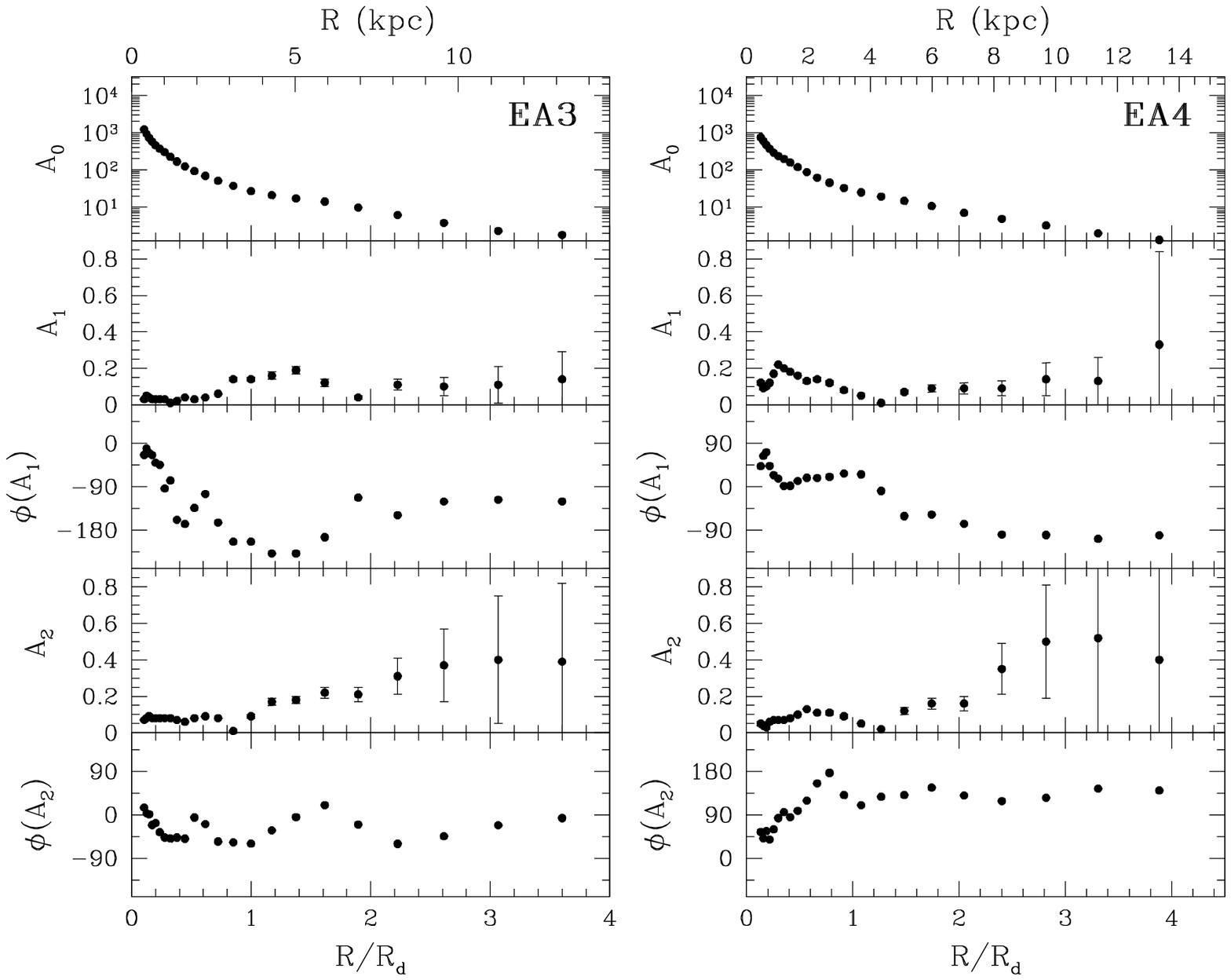}
\caption{Amplitudes and phases of each Fourier component in the F702W
  images. Radius is plotted in disk scale units and kpc.
  $A_m$ and $\phi(A_m)$ are the m-th order Fourier amplitudes and phase 
  angles, respectively. The zeroth order component $A_0$ has no phase and 
  reflects the mean flux at the given radius. The first order component $A_1$ 
  denotes the lopsidedness of the galaxy at the given radius.
\label{fig:fourier}}
\end{figure}

\begin{figure}
\figurenum{6}
\epsscale{0.7}
\plotone{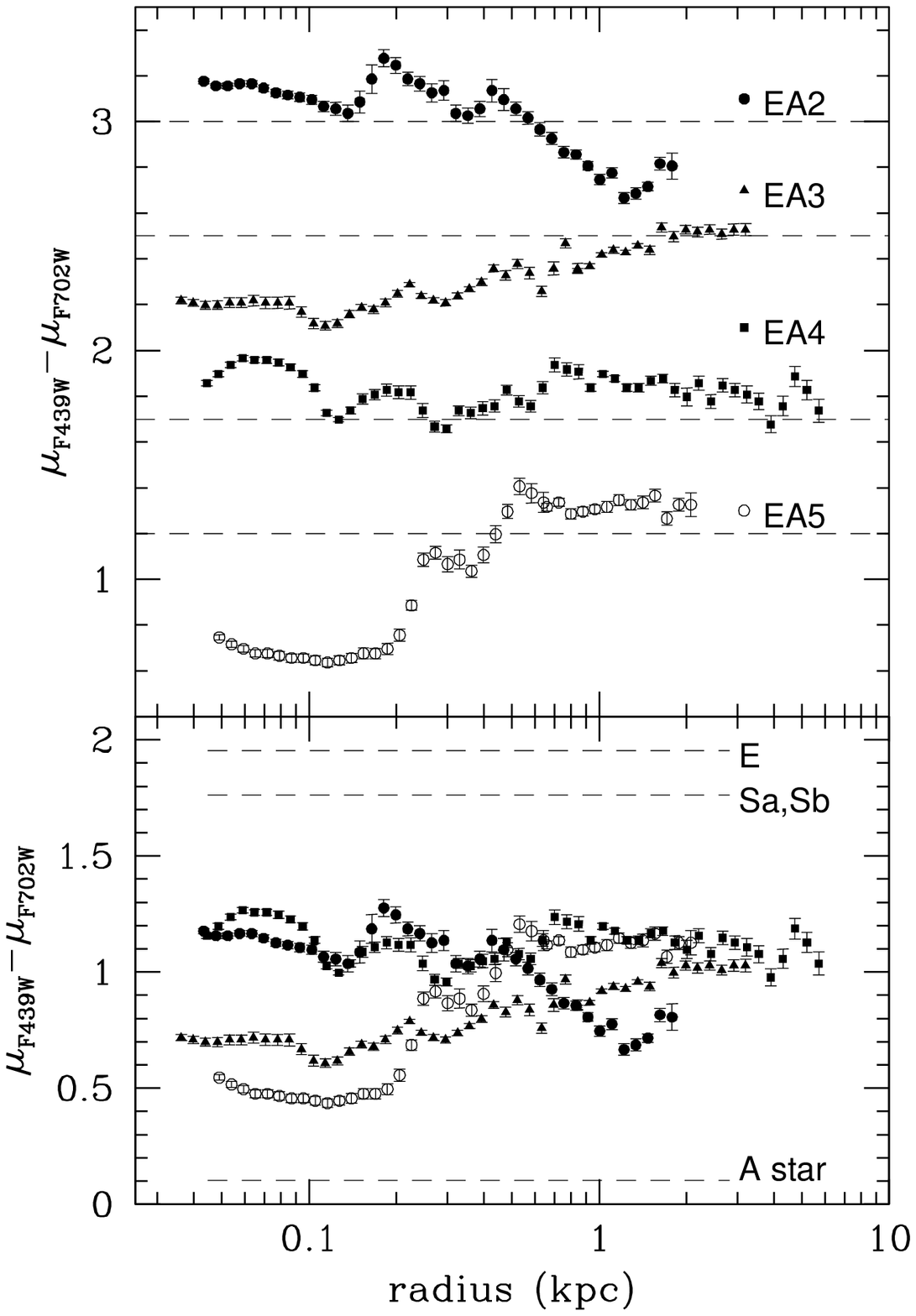}
\caption{$\mathrm{(F702W-F439W)}$ color profiles of four E+A galaxies. 
  The zero point in the color axis is arbitrary in the upper panel. 
  The dashed $\mathrm{(F702W-F439W)}=1.0$ lines for each color
  profile are shown for reference. The dashed lines in the lower 
  panel represent the typical colors of elliptical galaxies, spiral galaxies, 
  and A stars. Notice the diversity of the color gradients  
  within 1 kpc, which is reminiscent of their morphological 
  diversity.\label{fig:color}}
\end{figure}

\begin{figure}
\figurenum{7}
\epsscale{1.0}
\plotone{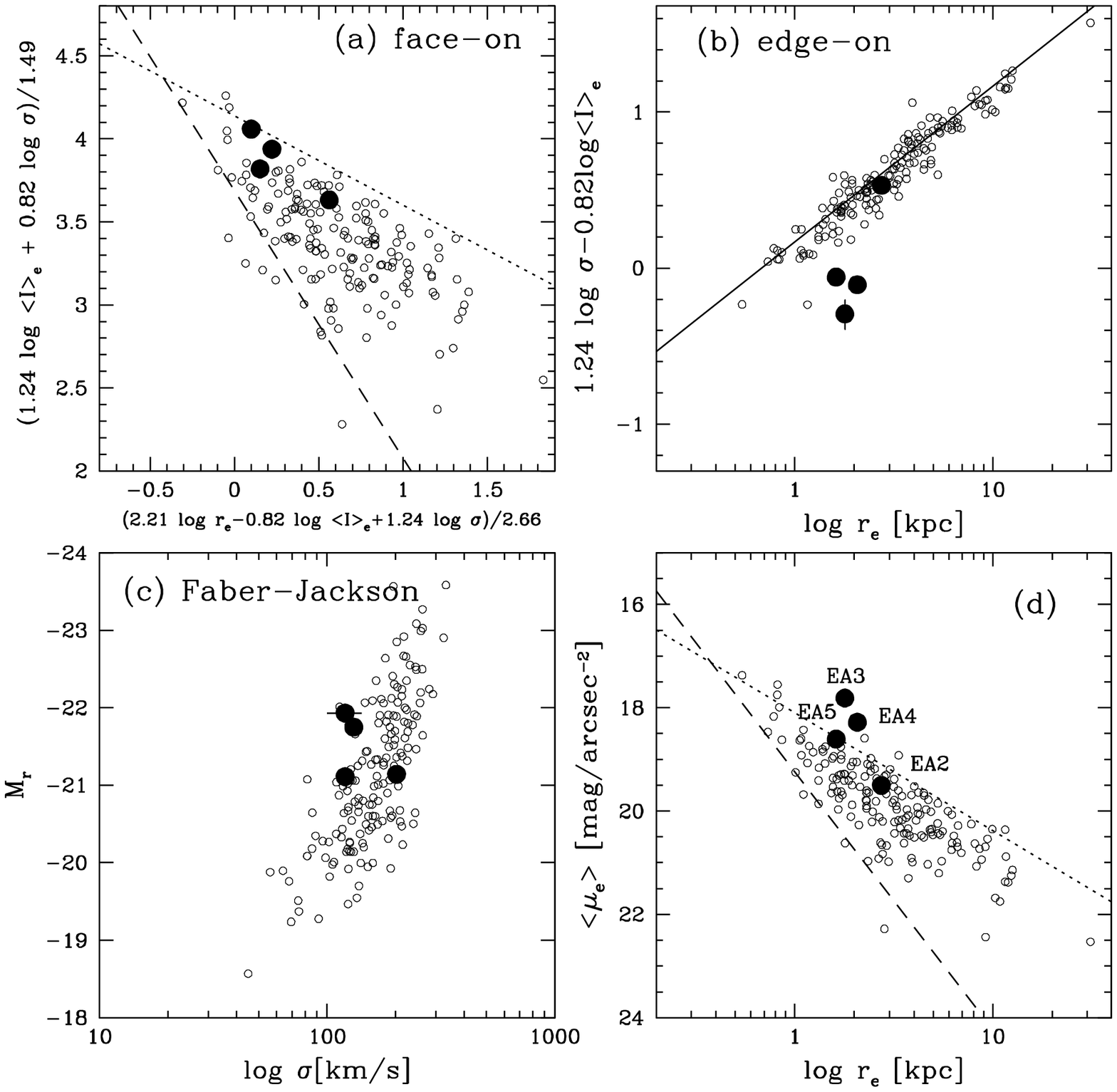}
\caption{Relative position of four E+A galaxies in the Fundamental Plane (FP). 
  The small open circles are data from Jorgensen et al (1996).
  (\emph{a}) Face-on view of the FP as in Jorgensen et al (1996).
  The dashed line indicates the boundary set by the limiting magnitude, but
  the upper dotted line is not caused by a selection effect.
  (\emph{b}) Edge-on view of the FP in the longest direction 
  (dashed line in panel (\emph{a})) of distribution. 
  (\emph{c}) The Faber-Jackson relation.
  Our galaxies occupy a small region in the parameter
  space.  (\emph{d}) The $r_e-\mu_e$ correlations of early type
  galaxies and the four E+As. Notice the deviation of EA3 and EA4 from
  the average relation ($\sim 0.5 - 1.0$ magnitudes brighter than E/S0
  galaxies within one effective radius).\label{fig:FP}}
\end{figure}

\begin{figure}
\figurenum{8}
\plotone{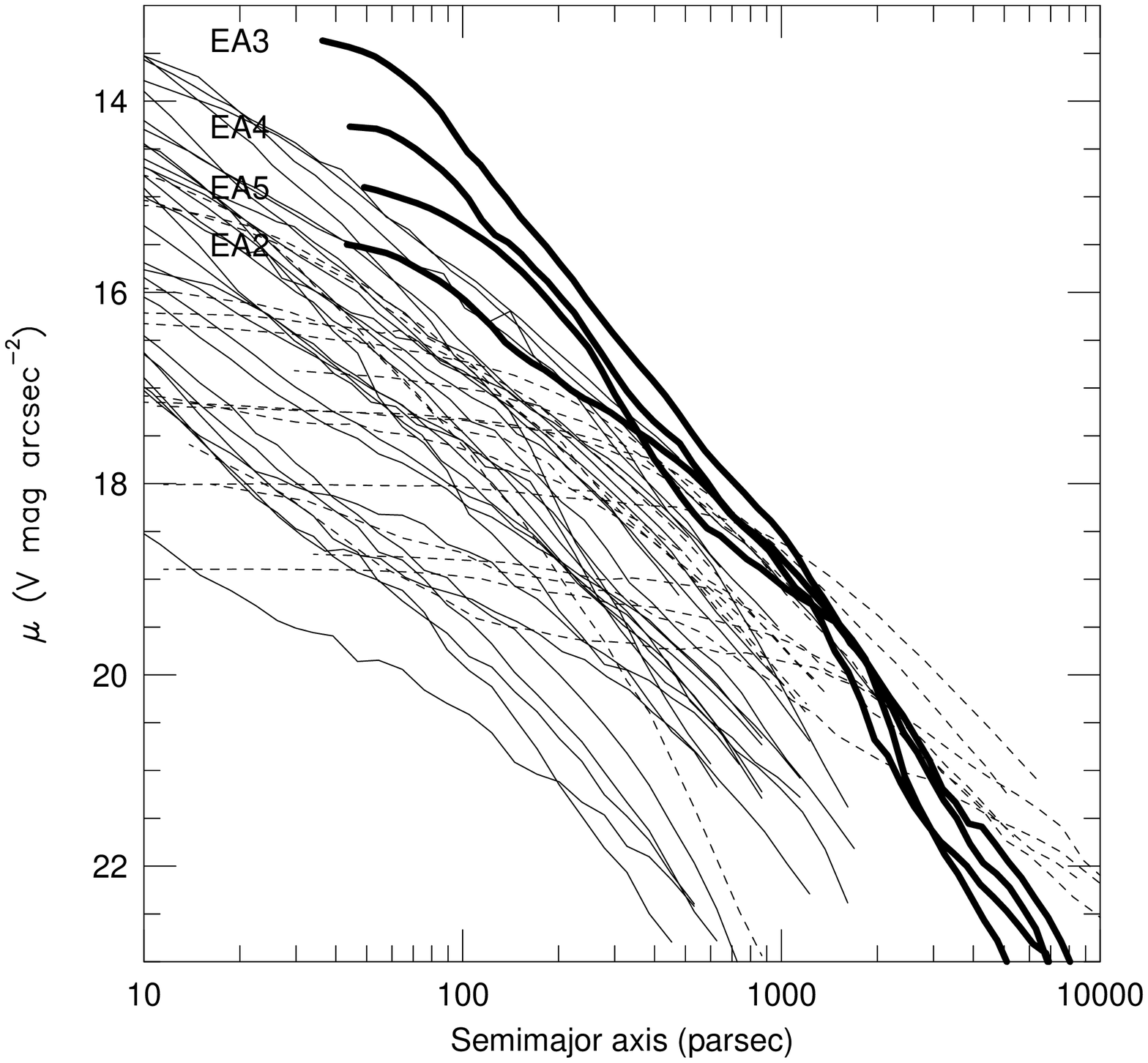}
\caption{The F702W surface brightness plots of the sample E+A galaxies versus
  those of normal early-type galaxies \citep{f97}. 
  (The present figure is an adaptation of Faber et al.'s Figure 1.)
  The E+As are matched to the \citet{f97} scale by assuming $V-R=0.5$ mag 
  and adopting $H_0=70$. Power-law galaxies are plotted as solid lines; 
  core galaxies are plotted as dashed lines.
  E+A galaxies (thick solid lines) have profiles like those of normal power-law 
  early-type galaxies, but scaled-up in surface brightness.
\label{fig:nuker}}
\end{figure}

\begin{figure}
\figurenum{9}
\plotone{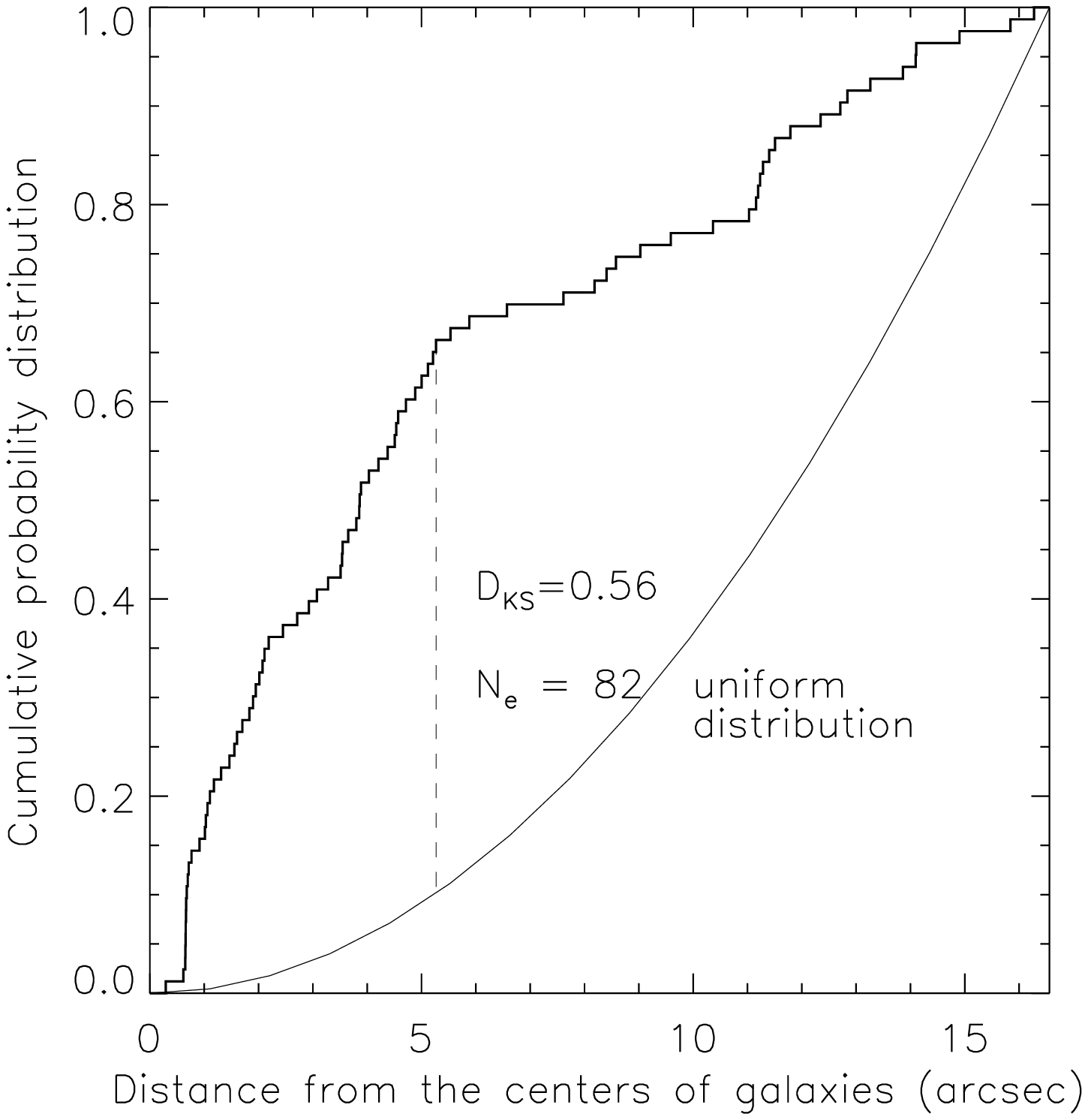}
\caption{Cumulative distribution of the angular distances of the candidate clusters
  from the center of the E+As. Thin solid line represents a uniform (or random) 
  distribution, $P(<r) \varpropto r^2$.
\label{fig:kstest}}
\end{figure}

\begin{figure}
\figurenum{10}
\epsscale{0.9}
\plotone{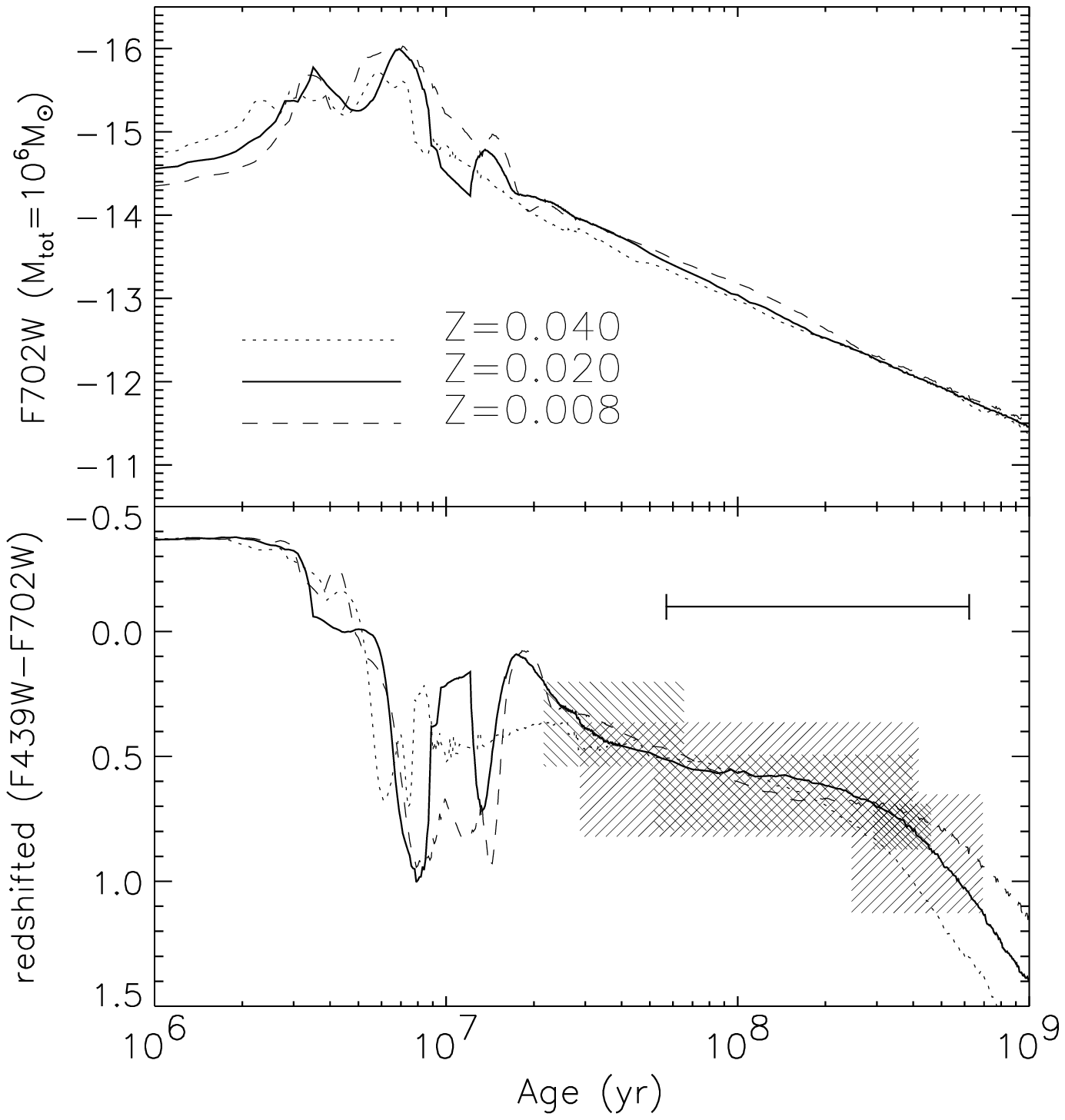}
\caption{The evolution of the redshifted (F439W-F702W) color and F702W magnitude of a
  simple stellar population calculated from the models of Starburst99
  \citep{starburst99}. Evolution is shown for a single starburst,
  Salpeter initial mass function \citep{Salpeter}, and three different metallicities. 
  The shaded regions represent the range of colors observed and 
  the corresponding range of ages for each cluster candidate in EA1.
  These candidates are identified in both the F702W and F439W image.
  The horizontal line in the bottom panel indicates the 95\% confidence
  level of the estimated single age of the clusters --- 
  the time elapsed since the starburst.
\label{fig:evolution}}
\end{figure}

\begin{figure}
\figurenum{11}
\plotone{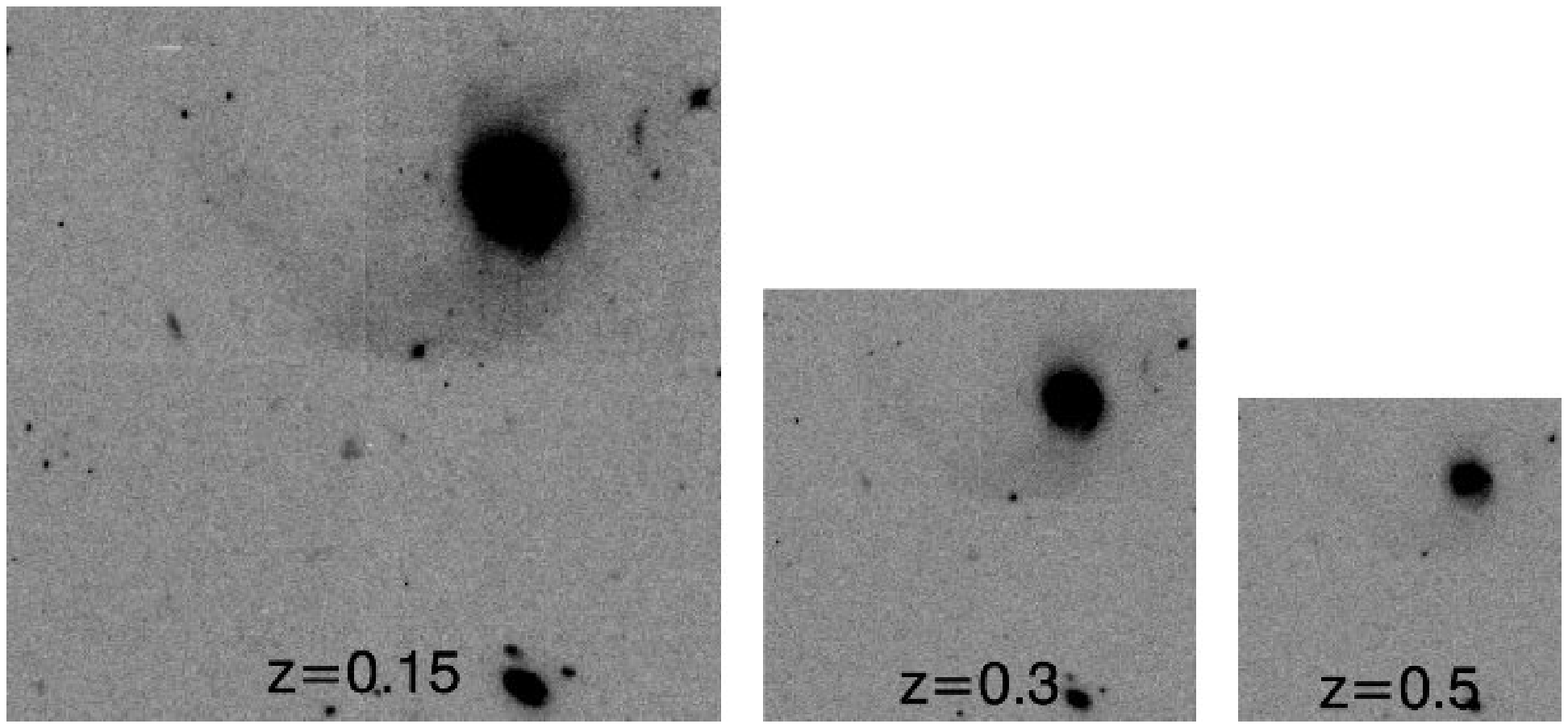}
\caption{
  Artificially redshifted images of EA3. We rebin the F702W band image of EA3
  and fade it artificially to mimic what it would look like at higher redshifts.
  The rebined images are convolved with the PSF, and then the sky background 
  and noise proportional to the exposure time are added to the redshifted images.
\label{fig:highz}}
\end{figure}

\end{document}